\documentclass[journal,10pt]{IEEEtran}

\IEEEoverridecommandlockouts

\usepackage{citesort}
\usepackage{epsfig}
\usepackage{graphicx}
\usepackage{amsfonts}
\usepackage{amsmath}
\usepackage{amssymb}
\usepackage{multicol}
\usepackage{psfrag}
\usepackage{subfigure}
\usepackage{stfloats}
\usepackage{footnote}
\usepackage{array}
\usepackage{algorithm}
\usepackage{algorithmic}
\usepackage{color}

\newtheorem{theorem}{Theorem}%[section]
\newtheorem{lemma}{Lemma}
\allowdisplaybreaks

\newtheorem{corollary}{Corollary}
\newtheorem{proposition}{Proposition}

\begin{document}

\title{Truncated Channel Inversion Power Control to Enable One-Way URLLC with Imperfect Channel Reciprocity}

\author{Chunhui~Li,~\IEEEmembership{Member,~IEEE,}
Shihao~Yan,~\IEEEmembership{Member,~IEEE,}\\
Nan~Yang,~\IEEEmembership{Senior Member,~IEEE,} and
Xiangyun~Zhou,~\IEEEmembership{Senior Member,~IEEE}

\thanks{This work was supported by the Australian Research Council under Discovery Project Grant DP180104062.}
\thanks{Part of this work has been presented to IEEE Globecom 2019 \cite{Li2019GlobeCOMW}.}
\thanks{C. Li, N. Yang, and X. Zhou are with the School of Engineering, The Australian National University, Canberra, ACT 2600, Australia (Emails: \{chunhui.li, nan.yang, xiangyun.zhou\}@anu.edu.au).}
\thanks{S. Yan is with the School of Science, Edith Cowan University, Perth, WA 6027, Australia (Email: s.yan@ecu.edu.au).}}

%\markboth{Submitted to IEEE Transactions on Communications}%
%{Li \MakeLowercase{\textit{et al.}}: Truncated Channel Inversion Power Control to Enable One-Way URLLC with Imperfect Channel Reciprocity}

\maketitle

\begin{abstract}
We propose to use channel inversion power control (CIPC) to achieve one-way ultra-reliable and lowlatency communications (URLLC), where only the transmission in one direction requires ultra reliability and low latency. Based on channel reciprocity, our proposed CIPC schemes guarantee the power of received signal that is used to decode the information to be a constant value $Q$, by varying the transmit signal and power, which relaxes the assumption of knowing channel state information (CSI) at the user. Thus, the CIPC schemes eliminate the overhead of CSI feedback, reduce communication latency, and explore the benefits of multiple antennas to significantly improve transmission reliability.
We derive analytical expressions for the packet loss probability of the proposed CIPC schemes, based on which we determine a closed interval and a convex set for optimizing $Q$ in CIPC with imperfect and perfect channel reciprocity, respectively. Our results show that CIPC is an effective means to achieve one-way URLLC.
The tradeoff among reliability, latency, and required resources (e.g., transmit antennas) is further revealed, which provides novel principles for designing one-way URLLC systems.
\end{abstract}


\IEEEpeerreviewmaketitle

\section{Introduction}

Ultra-reliable and low-latency communications (URLLC) is envisioned to support mission critical applications with stringent requirements of latency and reliability, e.g., industrial automation and autonomous vehicles. Specifically, in URLLC scenarios, the end-to-end delay and the decoding error probability are on the order of $1$ ms and $10^{-7}$, respectively~\cite{Popovski2018NETW,Bennis2018PROC,Li2018PROCEDIEEE,Chen2018MCOM,She2020PROC}. Due to its significance and potential, the key parameters for URLLC in terms of delay, reliability, packet size, network architecture and topology have been identified in \cite{Bennis2018PROC,Popovski2018ACCESS,She2020PROC}. Moreover, fundamental advancements in URLLC have been made in the context of various applications and techniques, e.g., mobile edge computing \cite{Dong2019TWC,Elbamby2019PROC}, non-orthogonal multiple access\cite{Yu2018COML,Sun2018TWC,Xiao2019JSAC}, physical layer security \cite{Ri2019WCOM,Wang2019TWC}, cooperative relaying \cite{Gu2018WCOML,Hu2019JSAC}, wireless energy transfer \cite{Lopez2018JS,Hu2019JSAC}, network slicing \cite{Popovski2018ACCESS,Kalor2018TINDINF}, industrial control \cite{Luvisotto2017TINDINF,Zhan2019TINDINF}, and radio resource management \cite{She2017MCOM,She2018TWC}.
%Specifically, Besides, the relaying-enabled URLLC networks have been first modelled and analyzed in \cite{Hu2018NETW}, which indicates the potential benefits of applying relaying in URLLC.
Recently, a novel architecture that applies deep learning with wireless edge intelligence for URLLC has been proposed in \cite{Gu2020DRL,She2020PROC}, which aims to provide practical guidelines into latency reduction in URLLC applications.

Considering the low-latency constraint, the coding blocklength (i.e., channel uses or packet size) is required to be as short as possible in URLLC applications~\cite{Johansson2015ICCW,Yilmaz2015ICCW}. Despite this, it is a huge challenge to satisfy the quality-of-service (QoS) requirements (i.e., the ultra-reliable and low-latency requirements) when the coding blocklength becomes short and limited in practice. Specifically, the decoding error probability is no longer negligible for finite blocklength and accurate channel state information (CSI) is hard to achieve in wireless networks within a short time period. Existing studies, aiming at ensuring the QoS requirements of URLLC in the finite blocklength regime, mainly assumed that the CSI is available or can be accurately estimated by using negligible channel uses. For instance, radio resource management in the finite blocklength regime was investigated to satisfy QoS requirements with signaling overhead for downlink transmission via cross-layer resource allocation in \cite{She2018TWC}, and for short packet delivery via joint uplink and downlink optimization in \cite{She2018TCOM}. In \cite{Hu2018TWC}, the optimal power allocation was studied for QoS-constrained downlink multi-user networks with different types of data arrival. In the aforementioned studies, the cost of channel estimation to satisfy QoS requirements was ignored by adopting the assumption that CSI is \emph{a prior} available or estimated by using negligible resources. Meanwhile, the impact of channel estimation cost on transmitting short packets in the finite blocklength regime was examined in the literature (e.g., \cite{Durisi2016TOC,Li2019TVT}), which revealed that such cost can be dominant and significantly affects the achievable reliability, especially when the low-latency requirement is very stringent. Therefore, how to significantly reduce or avoid the channel estimation overhead becomes an urgent and challenging research problem in URLLC.

Following the aforementioned discussions, we note that traditional channel estimation with feedback may cost extra signalling overhead, which results in non-negligible transmission and processing delay. As such, the traditional channel estimation may not work properly in some URLLC services. Against this background, traditional channel estimation needs to be revisited or modified with novel design in order to fulfill the strict requirements of URLLC.
%\red{When channel reciprocity holds, channel inversion power control (CIPC) can be used for wireless communications, while eliminating the conventional requirement that a user needs to know the CSI for decoding \cite{ElSawy2014TCOM,ElSawy2014TWC}. This is due to the fact that a base station (BS) can use CIPC to vary its transmit signal and power, in order to ensure that the signal power that \red{is} used to decode the information at the user is a constant value, which is \emph{a prior} agreed between the BS and the user.
%We note that CIPC requires the CSI being available at the BS, but avoids the cost of feeding back the estimated CSI from the BS to the user.}
When channel reciprocity holds, channel inversion power control (CIPC) can be used for wireless communications, while eliminating the conventional requirement that a user needs to know the CSI for decoding information \cite{ElSawy2014TCOM,ElSawy2014TWC}. This is due to the fact that a base station (BS) can use CIPC to adjust the magnitude and phase of the channel by varying the transmit power with a proper precoding signal. As a consequence, the effective channel is a constant value, which is \emph{a prior} agreed between the BS and the user.
This property leads to that CIPC may serve as a key enabler of one-way URLLC in future wireless networks. Notably, one-way URLLC has a wide range of applications.
For example, in a smart factory the communication from a central controller to an actuator that delivers an urgent message triggered by the reception and processing of the uplink information requires one-way URLLC, while the communication on the other way (mainly delivering periodically measurements or updates) may not require URLLC. Similar application scenarios can also be found in digital medical systems and industrial Internet of Things.

Although CIPC has been studied in different communication scenarios (e.g., \cite{ElSawy2014TCOM,ElSawy2014TWC,Hu2019TVT}), its performance and the associated optimization of the agreed constant power have never been investigated in the context of URLLC.
It is worthwhile to mention that Shannon’s capacity is used in previous works to characterize the maximal achievable rate in the infinite blocklength regime, which is jointly convex in bandwidth and transmit power. However, considering finite blocklength in URLLC, the maximal achievable rate is neither convex nor concave with respect to radio resources (i.e., bandwidth and transmit power) \cite{Xu2016TWC,Sun2019TWC}, which leads to non-convex QoS constraints in optimizing resource allocation for URLLC. It is very challenging to address such non-convex optimization problem in URLLC.
Besides, the maximal achievable rate in the finite blocklength regime is lower than Shannon’s capacity, which leads to the fact that the latency and reliability will be underestimated if using Shannon’s capacity in optimizing resource allocation for URLLC \cite{Schiessl2015MSWiM}. The above issues motivate this work to tackle the feasibility and the optimal design of using CIPC to achieve the one-way URLLC and establish the fundamental limit of one-way URLLC achieved by CIPC.
%This motivates this work to tackle the feasiblity and the optimal design of using CIPC to achieve the one-way URLLC and establish the fundamental limit of one-way URLLC achieved by CIPC.

%In one-way URLLC, only the transmission in one direction requires ultra reliability and low latency, while the transmission in the other direction does not have strict requirements on delay and latency. This makes it possible for a user to periodically broadcast pilots such that the BS can estimate CSI. The BS has to take the estimated uplink CSI as the downlink CSI for the one-way URLLC. Thus, channel reciprocity is required in this context, but may not be perfect in practice due to the challenging channel hardening issues \cite{Chen2018TCOM,Ngo2017TWC,Mi2017TCOM}.

In addition, channel reciprocity may not be perfect in practice due to the challenging channel hardening issues \cite{Chen2018TCOM,Ngo2017TWC,Mi2017TCOM}. As such, we consider both the perfect and imperfect channel reciprocity in this work and examine the impact of the imperfectness on the performance of CIPC in one-way URLLC. In CIPC, the power at the BS needs to approach infinity to ensure the power of the received signal that is used to decode the useful information being a constant for some low-quality channel realizations. As such, another key factor that limits the performance of CIPC is the maximum transmit power constraint, which determines when the BS has to suspend its transmission. Therefore, in this work we also consider the maximum transmit power constraint for the considered CIPC, which leads to the truncated CIPC scheme. Our contributions in this work are summarized as follows:
\begin{itemize}
\item
We develop a fundamental framework for using the truncated CIPC scheme to achieve one-way URLLC. To this end, we first derive a new expression for the packet loss probability $P_{\epsilon}$ of this scheme, which is determined by both the decoding error probability caused by the finite blocklength $T$ and the transmission probability enforced by the maximum transmit power $P_{\mathrm{max}}$. We then derive an approximated but easy-to-calculate expression for $P_{\epsilon}$ as a function of the channel reciprocity characterised by the parameter $\phi$, based on which we explicitly determine a closed interval for the optimal constant value $Q$ to minimize $P_{\epsilon}$, where $Q$ is the received signal power \textit{a priori} agreed between the BS and the user.
\item
We analyze the performance of the conventional CIPC scheme in the context of one-way URLLC, which is the special case of the truncated CIPC scheme with $P_{\mathrm{max}} \rightarrow \infty$. The packet loss probability $P_{\epsilon}$ of the conventional CIPC scheme converts into the decoding error probability caused by the finite blocklength $T$, which enables us to derive a closed-form expression (as an explicit expression of the blocklength $T$ and the channel reciprocity parameter $\phi$) to approximate $P_{\epsilon}$. The packet loss probability of the conventional CIPC scheme offers an upper bound on the performance of the truncated CIPC scheme. Thus, the closed-form expression significantly facilitates us to examine the performance limit of one-way URLLC achieved by CIPC with imperfect channel reciprocity (i.e., $0<\phi<1$).
\item
%{(We consider two scenarios: (1) conventional CIPC scheme, i.e., $P_{\mathrm{max}} \rightarrow \infty$ (2) proposed truncated CIPC scheme, i.e., $P_{\mathrm{max}}$ is pre-determined, which includes two cases (a) perfect channel reciprocity (i.e., $\phi=1$) and (b) imperfect channel reciprocity (i.e., $\phi \neq 1$).)}
We derive the packet loss probability for the truncated CIPC scheme with perfect channel reciprocity (i.e., $\phi=1$), denoted as $P^{\phi=1}_{\epsilon}$, which is not a special case of that for the truncated CIPC scheme with imperfect channel reciprocity. We analyze the convexity of $P_{\epsilon}$ with respect to (w.r.t.) $Q$ and establish a convex set for optimizing the value of $Q$ in the truncated CIPC scheme with perfect channel reciprocity. Our examination draws novel design guidelines for achieving one-way URLLC with the CIPC scheme. For instance, the maximum transmit power or number of transmit antennas can be explicitly determined using our analysis to perform a fixed-rate transmission with specific requirements on reliability and latency.
%\blue{Furthermore, our results show that increasing the number of transmit antennas continuously improves the performance of one-way URLLC with CIPC, which is different from the URLLC that is based on traditional channel estimation and feedback mechanisms.} {(This sentence needs to be revised.)}
\end{itemize}

%\emph{Notations:} Vectors and matrices are denoted by lower-case and upper-case boldface symbols, respectively. Given a complex number $z$, $|z|$ denotes the modulus of $z$. Given a complex vector $\mathbf{x}$, $\|\mathbf{x}\|$ denotes the Euclidean norm, $\mathbf{x}^{\ast}$ denotes the conjugate of $\mathbf{x}$, and $\mathbf{x}^{\mathcal{T}}$ denotes the transpose of $\mathbf{x}$. $\mathbf{I}_{N_{\mathrm{t}}}$ denotes an $N_{\mathrm{t}}\times N_{\mathrm{t}}$ identity matrix and $\mathbb{E}[\cdot]$ denotes expectation. $\mathcal{CN}\left(\mu,\nu\right)$ denotes the circularly symmetric complex Gaussian distribution with the mean of $\mu$ and variance of $\nu$.

\section{System Model}\label{system_model}

In this section, we first detail our considered scenario of one-way URLLC together with the adopted assumptions. Then, we present our proposed CIPC scheme and its associated performance metric, i.e., the packet loss probability.

\begin{figure}[!t]
\begin{center}
    \includegraphics[width=0.8\columnwidth]{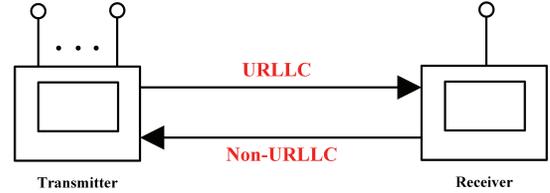}
    \caption{Illustration of the considered one-way URLLC scenario.}\label{system_model_figure}%\vspace{-2em}
\end{center}
\end{figure}

As shown in Fig.~\ref{system_model_figure}, in this work we consider a downlink one-way URLLC scenario in a time division duplex (TDD) multiple-input single-output (MISO) communication system, where an $N_{\mathrm{t}}$-antenna BS sends an urgent message triggered by the reception and processing of the uplink information to a single-antenna user with the stringent requirement of latency and reliability.
Specifically, we consider the URLLC occurring in the downlink, while the uplink transmission is non-URLLC. The user periodically sends regular information in the uplink to the BS. The regular information consists of both pilots and useful information in order to enable BS to jointly perform channel estimation and information decoding. Upon reception and processing of the uplink information, BS decides whether or not to send an urgent message to the user in the downlink. Hence, the downlink URLLC is actually triggered by the reception and processing of the uplink information. This means that the downlink URLLC does not happen at any arbitrary point in time, but it can only happen right after the reception and processing of a non-URLLC communication in the uplink. This is an practically important scenario. For example, in the context of a smart factory, a central controller controls multiple moving robots and actuators in a factory. To avoid the collisions of moving robots, accurate downlink controls with low latency and high reliability are needed at all the time. The control decisions are mainly based on the feedback information not only from robots and actuators but also from additional monitoring sensors. In order to achieve the near-perfect CSI at the controller, many advanced technologies can be utilized to enhance the accuracy of CSI estimation. It could deploy intelligent reflecting surface and no blind spot monitoring camera in a factory to provide additional information for the controller to make a decision whether a triggering condition of a downlink URLLC is satisfied or not.
The processing task of analyzing the network could be completed by well-trained deep neural networks with proper online fine-tuning. It is noted that the processing delay could be one transmission time interval (TTI) ($0.125\sim 1$~ms) in 5G New Radio (NR) \cite{Gu2020DRL}. As such, if the controller decides to immediately send out an urgent control message to the user, this downlink URLLC happens with very minimal time gap from the uplink communication that just happened. Therefore, it is reasonable to assume that the channel has not changed significantly from the uplink communication to the downlink communication. It is followed that, the controller can use the estimated channel information to perform downlink communication with channel inversion power control.

%The processing task of analyzing the network could be completed by well-trained deep neural networks with proper online fine-tuning. It is noted that the processing delay could be one transmission time interval (TTI) ($0.125\sim 1$~ms) in 5G New Radio (NR) \cite{Gu2020DRL}. As such, if the BS decides to immediately send out an urgent control message to the user, this downlink URLLC happens with very minimal time gap from the uplink communication that just happened. Therefore, it is reasonable to assume that the channel has not changed significantly from the uplink communication to the downlink communication. It is followed that, the BS can use the estimated channel information to perform downlink communication with channel inversion power control.

We denote $\mathbf{h}_{\mathrm{u}}$ as the $N_{\mathrm{t}} \times 1$ uplink channel vector from the user to the BS and denote $\mathbf{h}_{\mathrm{d}}$ as the $1 \times N_{\mathrm{t}}$ downlink channel vector from the BS to the user. Specifically, the downlink transmission considered in this work requires URLLC, i.e., the downlink transmission with a high reliability requirement needs to be performed within a finite blocklength $T$ (or equivalently, $T$ channel uses). All the channels are subject to independent quasi-static Rayleigh fading such that the entries of each channel vector are assumed to be independent and identically distributed (i.i.d.) circularly symmetric complex Gaussian random variables with zero mean and unit variance, i.e., $\mathbf{h}_{\mathrm{d}}\sim\mathcal{CN} \left(0,\mathbf{I}_{N_{\mathrm{t}}}\right)$ and $\mathbf{h}_{\mathrm{u}}\sim\mathcal{CN} \left(0,\mathbf{I}_{N_{\mathrm{t}}}\right)$.

Considering the imperfect channel reciprocity, the downlink channel vector can be expressed as a function of the uplink channel vector, given by \cite{Mi2017TCOM,Li2018ICCW}
\begin{align}\label{eq:dowlink_channel}
\mathbf{h}_{\mathrm{d}}=\sqrt{\phi}\mathbf{h}_{\mathrm{u}}^{\mathcal{T}}
+\sqrt{1-\phi}\mathbf{e}^{\mathcal{T}},
\end{align}
where $\phi$ is defined as the channel reciprocity coefficient between the uplink and downlink channels, and $\mathbf{e}$ is the $N_\mathrm{t} \times 1$ vector that reflects the uncertain part of $\mathbf{h}_{\mathrm{u}}$. The entries of $\mathbf{e}$ are i.i.d. and each of them follows $\mathcal{CN}\left(0,1\right)$. We note that $\mathbf{e}$ is independent of $\mathbf{h}_{\mathrm{u}}$. The adoption of the channel reciprocity model in \eqref{eq:dowlink_channel} is motivated by the fact that channel reciprocity requires appropriate hardware calibrations to compensate for the unknown amplitude scaling and phase shift between the downlink and uplink channels in practice~\cite{Kaltenberger2010Proc}. The value of $\phi$ quantifies the level of channel reciprocity, where $0\leq\phi\leq1$. In practical scenarios, the level of channel reciprocity is determined by the uplink channel estimation error and the frequency offset between the BS and the user~\cite{Kaltenberger2010Proc}. Specifically, $\phi=1$ indicates that the perfect channel reciprocity is achieved such that the downlink channel is exactly the same as the uplink channel. As $\phi$ decreases, the channel reciprocity decreases. When $\phi=0$, the channel reciprocity does not exist such that the downlink channel is independent of the uplink channel.

%Similar application scenarios can also be found in digital medical systems and industrial Internet of Things.

We assume that the BS knows $\mathbf{h}_{\mathrm{u}}$ perfectly. This is due to the fact that the uplink transmission does not have strict requirement on latency, which makes it possible for the user to periodically broadcast pilots for allowing the BS to estimate $\mathbf{h}_{\mathrm{u}}$.
Even if the estimation is not perfect, the estimation error can be incorporated into the channel reciprocity coefficient $\phi$ in \eqref{eq:dowlink_channel}. In order to enable a user (e.g., a vehicle) to decode the information without knowing CSI, the CIPC scheme is used at the BS based on $\mathbf{h}_{\mathrm{u}}$, which will be detailed in the following subsection. The CIPC scheme can significantly reduce the communication latency and improve transmission reliability to meet the requirements of URLLC. This is due to the fact that the CIPC scheme saves the signaling overhead used to feed back the estimated $\mathbf{h}_{\mathrm{u}}$ from BS to a user and allows all the downlink channel uses being available for data transmission when urgent information transmission is on demand.

We note that the considered one-way URLLC model could be generalized to two-way URLLC cases in future works. However, there exist multiple challenges in achieving the two-way URLLC. For example, in TDD systems, without full-duplex techniques, the transmitter and the receiver cannot transmit and receive the message simultaneously, which increases the latency. We also note that, if the full-duplex techniques are used, the self-interference may limit the communication reliability and thus cannot obtain URLLC. In addition, channel estimation has to be conducted at one transceiver in TDD systems, which may cost extra latency and lead difficulties in achieving two-way URLLC. The main motivation of our this work, i.e., considering CIPC, is to remove the latency caused by channel estimation. Motivated by our work, actually one method to achieve two-way URLLC is to use CIPC at two different frequency channels, i.e., one for uplink transmission and the other one for downlink transmission. This means that the two-way URLLC can be achieved at the cost of extra frequency bandwidth with the aid of our proposed CIPC.

\subsection{Channel Inversion Power Control}

In this work, the CIPC scheme is used at the BS to enable the user to decode the received signals without knowing $\mathbf{h}_{\mathrm{d}}$. The received signal in one channel use is given by
\begin{align}\label{eq:recei_signal_original}
y=\sqrt{P_{\mathrm{a}}}\mathbf{h}_{\mathrm{d}}\mathbf{x}+w,
\end{align}
%%%%%%%%%%%
where $w$ is the additive white Gaussian noise (AWGN) at the user with zero mean and variance $\sigma_{w}^{2}$, $\mathbf{x}$ is the transmitted signal which is subject to the average power constraint, i.e., $\mathbb{E}\left[\|\mathbf{x}\|^{2}\right]=1$, and $P_{\mathrm{a}}$ is the transmit power. Following the imperfect channel reciprocity model given in \eqref{eq:dowlink_channel}, the received signal in \eqref{eq:recei_signal_original} can be rewritten as
%%%%%%%%%%
\begin{align}\label{eq:recei_signal}
y=\sqrt{P_{\mathrm{a}}\phi}\mathbf{h}_{\mathrm{u}}^{\mathcal{T}}\mathbf{x}
+\sqrt{P_{\mathrm{a}}\left(1-\phi\right)}\mathbf{e}^{\mathcal{T}}\mathbf{x}+w.
\end{align}
{It is noted that $\mathbf{e}$ changes for each fading block. In other words, $\mathbf{e}$ is fixed within the considered blocklenth (i.e., channel uses or symbols). While $\mathbf{x}$ is different for each channel uses within the blocklength. Generally, the distribution of $\mathbf{e}$ cannot be determined. Following \cite{Hassibi2003JIT,Li2019TVT,Sun2020TCOM}, the worst-case scenario for the decoding process at the user could be considered, where $\sqrt{P_{\mathrm{a}}\left(1-\phi\right)}\mathbf{e}^{\mathcal{T}}\mathbf{x}+w$ could be approximated as a zero-mean Gaussian noise. Under this approximation, the packet loss probability could work as an upper bound.}

In order to counteract the impact of downlink channel phase at the user, the transmitted signal $\mathbf{x}$ is written as
\begin{align}\label{eq:trans_signal}
\mathbf{x}=\frac{\mathbf{h}^{\ast}_{\mathrm{u}}}{\|\mathbf{h}_{\mathrm{u}}\|}u,
\end{align}
where $u$ is the information signal transmitted from the BS to the user.
Following \eqref{eq:recei_signal} and \eqref{eq:trans_signal}, the signal-to-interference-plus-noise ratio (SINR) at the user can be written as
\begin{align}\label{eq:SINR}
\gamma&=\frac{P_{\mathrm{a}}\phi\|\mathbf{h}_{\mathrm{u}}\|^{2}}
{P_{\mathrm{a}}\left(1-\phi\right)\frac{{\left|\mathbf{e}^{\mathcal{T}}\mathbf{h}_{\mathrm{u}}^{\ast} \right|}^2}{{\left\|\mathbf{h}_{\mathrm{u}}\right\|}^2} +\sigma_{w}^{2}}.
\end{align}

In order to counteract the impact of downlink channel gain at the user, the BS adopts the CIPC scheme to vary its transmit power as per $\|\mathbf{h}_{\mathrm{u}}\|$, given by
\begin{align}\label{constant_Q}
P_{\mathrm{a}}\|\mathbf{h}_{\mathrm{u}}\|^{2}=Q,
\end{align}
where $Q$ is a pre-determined constant value \textit{a priori} agreed between the BS and the user. Then, the SINR at the user in \eqref{eq:SINR} can be rewritten as
\begin{align}\label{eq:SINRTruncatedCIPC_new}
\gamma=\frac{\phi Q}{\left(1-\phi\right)\frac{Q\left|\mathbf{e}^{\mathcal{T}} \mathbf{h}_{\mathrm{u}}^{\ast}\right|^{2}}{\left\|\mathbf{h}_{\mathrm{u}}\right\|^{2}
\left\|\mathbf{h}_{\mathrm{u}}\right\|^{2}}+\sigma_{w}^{2}}.
\end{align}

Considering Rayleigh fading for $\mathbf{h}_{\mathrm{u}}$, as per \eqref{constant_Q} we see that the transmit power $P_{\mathrm{a}}$ may be infinite to guarantee $P_{\mathrm{a}}\|\mathbf{h}_{\mathrm{u}}\|^{2}=Q$ for some realizations of $\mathbf{h}_{\mathrm{u}}$. Without loss of generality, in this work we consider a maximum transmit power constraint, denoted by $P_{\mathrm{max}}$ \cite{ElSawy2014TWC}. Specifically, the BS only transmits information to the user when the uplink channel gain, i.e., $\|\mathbf{h}_{\mathrm{u}}\|^{2}$, is greater than a specific value. Mathematically, the transmit power is given by
\begin{align}\label{eq:TransmitPowerAP}
P_{\mathrm{a}}&=
\begin{cases}
\frac{Q}{\|\mathbf{h}_{\mathrm{u}}\|^{2}} ,& \|\mathbf{h}_{\mathrm{u}}\|^{2} \geq \frac{Q}{P_{\mathrm{max}}}, \\
0 ,& \|\mathbf{h}_{\mathrm{u}}\|^{2} < \frac{Q}{P_{\mathrm{max}}}.
\end{cases}
\end{align}
%where $P_{\mathrm{max}}$ is the maximal transmit power.
In this work, we refer to the CIPC scheme with a finite $P_{\mathrm{max}}$ as the truncated CIPC scheme, where the transmit power is truncated at a specific value. As $P_{\mathrm{max}} \rightarrow \infty$, the truncated CIPC scheme converges to the conventional CIPC scheme, where transmission is always performed regardless of the channel quality. In this work, we first analyze the truncated CIPC scheme and then analyze the conventional CIPC scheme as a special case, which serves as a performance benchmark for the truncated CIPC scheme.

\subsection{Fixed-Rate Transmission and Packet Loss Probability}

As per \eqref{eq:TransmitPowerAP}, we see that the BS does not always transmit information to the user in the truncated CIPC scheme. The associated transmission probability, i.e., the probability that the BS sends information to the user, is given by
\begin{align}\label{trans_prob}
p_t(Q)=\mathrm{Pr}\left\{P_{\mathrm{a}}\leq P_{\mathrm{max}}\right\}.
\end{align}
We note that in the conventional CIPC scheme (i.e., where $P_{\mathrm{max}}\rightarrow\infty$), the transmission is always performed and thus, the transmission probability is one (i.e., $p_t(Q) = 1$).

In URLLC scenarios, the packet loss from the BS to the user is caused by not only the aforementioned transmission suspension induced by the considered maximum transmit power constraint. When the transmission is performed, the packet loss still occurs due to the non-zero decoding error probability for the finite blocklength \cite{Polyanskiy2010}.
%{In URLLC scenarios, the required quality of service (QoS) is normally predetermined in terms of a specific transmission rate or packet loss probability. In this work, we consider a fixed-rate transmission, where the information transmission rate $R$ from the BS to the user is predetermined and represents a certain requirement on QoS. We note that a varying-rate transmission can be considered in the context of URLLC, where the transmission rate can be adapted as per a predetermined packet loss probability (which represents a certain QoS in another form) and practical channel conditions.}
In URLLC scenarios, the required QoS is normally predetermined in terms of a specific transmission rate or packet loss probability. In this work, we consider a fixed-rate transmission, where the information transmission rate $R$ from the BS to the user is predetermined to meet a certain requirement on QoS. We note that varying-rate transmission can be considered in the context of URLLC, where the transmission rate can be adapted as per a predetermined packet loss probability (which represents a certain QoS in another form) and practical channel conditions.
%\footnote{We note that varying-rate transmission can be considered in the context of URLLC, where the transmission rate can be adapted as per a predetermined packet loss probability (which represents a certain QoS in another form) and practical channel conditions. The extension of our study to the varying-rate transmission scenario will be studied in future work.}.}
For fixed-rate transmission, based on a widely used asymptotic expression for the non-zero decoding error probability given in \cite{Polyanskiy2010}, the decoding error probability averaged over different channel realizations at the user can be approximated as
\begin{align}\label{eq:epsilon_def}
\epsilon&=\mathbb{E}_{\!\{\mathbf{h}_{\mathrm{u}}, \mathbf{e}\}\!}\left[\!f\left(\!\left(\log_{2}(1+\gamma)-R\right)\sqrt{\frac{T}{V}}\!\right)\!\right]
\notag\\
&=\mathbb{E}_{\!\{\mathbf{h}_{\mathrm{u}}, \mathbf{e}\}\!}\left[\!f\left(\frac{\sqrt{T}\left(\ln(1+\gamma)-R\ln2 \right)}{\sqrt{1-\left(1+\gamma\right)^{-2}}}\right)\!\right],
\end{align}
%\begin{align}
%\epsilon&=\mathbb{E}_{\{\mathbf{h}_{\mathrm{u}}, \mathbf{e}\}}\left[f\left(\frac{\log_{2}(1+\gamma)-R}{\sqrt{V/T}}\right)\right]\notag\\
%&=\mathbb{E}_{\{\mathbf{h}_{\mathrm{u}}, \mathbf{e}\}}\left[f\left(\frac{\sqrt{T}\left[\ln(1+\gamma)-R\ln2 \right]}{\sqrt{1-\frac{1}{(1+\gamma)^2}}}\right)\right],
%\end{align}
where we recall that $R$ is the information transmission rate, the SINR $\gamma$ is given in \eqref{eq:SINRTruncatedCIPC_new}, $V=(\log_{2}e)^2 \left(1-\left(1+\gamma\right)^{-2}\right)$ is the channel dispersion, and $f(\cdot)$ denotes the Gaussian $Q$-function with $f(x)=\frac{1}{\sqrt{2\pi}}\int_{x}^{\infty}e^{-\frac{t^2}{2}}dt$.
It is worth mentioning that $T$ is the blocklength and $T \geq 100$ is the condition for achieving effective approximation for the maximal achievable rate in the finite blocklength regime \cite{Polyanskiy2010}. We note that current practical codes have typical blocklength from 100 to 1000 for short-packet control information transmission, e.g., 168 in \cite{Durisi2016TOC}. Hence, $T \geq 100$ can be applied into practical scenarios.
{In addition, in this work, latency is characterized by the blocklength $T$ (i.e., the number of channel uses) instead of a specific equation describing the end-to-end latency. The main reason is the overall end-to-end latency in cellular networks is determined by the delays in radio access network (RAN), backhaul, core network, data center/cloud, and Internet server. It increases significantly with the network load and the transmission distance between the transmitter and the receiver. It is worth mentioning that communicating with the gateway of the core network toward the Internet takes a minimum of $39$ ms in Long Term Evolution (LTE) networks, where the latency in RAN contributes to a large proportion of the end-to-end latency (i.e., $10-20$ ms).
It is found that the assumption adopted commonly in 3GPP for a core/Internet latency component in LTE varies from $1$ to $20$ ms. Although the core network can be located close to the RAN to reduce the latency, the latency of $1$ ms cannot be achieved when considering the core network. For URLLC applications, the end-to-end latency mainly focuses on the latency in RAN. From the radio communication perspective, the latency in RAN is mainly contributed by the physical layer and media access control (MAC) layer. The strictly low latency in URLLC imposes an unprecedented restriction on the size of packets. Indeed, short packets have been recognized as the typical forms of the traffic generated by sensors and small mobile devices involved in mMTC and URLLC~\cite{Durisi2016TOC}. For example, in industrial manufacturing and control systems, measurements and control commands are of small size (e.g., 10 to 32 bytes)~\cite{Johansson2015ICCW,Yilmaz2015ICCW} and need to be communicated in real-time with ultra-high reliability. While in this work, we mainly focus on the physical layer. To achieve $1$ ms latency in RAN, a flexible frame structure in 5G New Radio (NR) is proposed for short transmission time interval (TTI) by using mini-slots (e.g., $2$ or $7$ OFDM symbols) instead of a fixed slot duration in LTE (durations of $14$ OFDM symbols). The slot length in 5G NR could be much shorter than $1$ ms, which depends on the selected numerology. Therefore, mini-slots can be used to address the latency requirements for URLLC scenarios from the physical layer perspective. Therefore, the size of blocklength is considered as a method to describe the latency in some extents. In other words, the latency is minimized in the physical layer when short blocklength is adopted. }

We note that a closed-form expression for \eqref{eq:epsilon_def} is intractable. Thus, we adopt the linear approximation given by
\begin{align}
f\left(\left(\log_{2}(1+\gamma)-R\right)\sqrt{\frac{T}{V}}\right)\approx\Omega(\gamma)
\end{align}
in this work, given by \cite{Makki2014WCL,Makki2016TOC}
\begin{align}
\Omega(\gamma) &\overset{\vartriangle}{=}
\begin{cases}
1 ,& 0 \leq \gamma \leq \alpha \\
\frac{1}{2}-\delta \left(\gamma -\gamma_{0}\right) ,& \alpha < \gamma < \beta \\
0 ,& \gamma \geq \beta,
\end{cases}
\end{align}
where $\delta=\frac{\sqrt{T}}{2\pi \sqrt{2^{2 R}-1}}$, $\gamma_{0}=2^{R}-1$, $\alpha=\gamma_{0}-\frac{1}{2\delta}$, and $\beta=\gamma_{0}+\frac{1}{2\delta}$. As such, the decoding error probability $\epsilon$ defined in \eqref{eq:epsilon_def} can be rewritten as
\begin{subequations}
\begin{align}\label{eq:esilon_approximation}
&\epsilon=\int_{0}^{\infty}f\left(\frac{\sqrt{T}\left(\ln(1+\gamma)-R\ln2 \right)}{\sqrt{1-\left(1+\gamma\right)^{-2}}}\right)f_{\gamma}(\gamma)d\gamma\\
&\approx\int_{0}^{\alpha}f_{\gamma}(\gamma)d\gamma
+\int_{\alpha}^{\beta}\left(\frac{1}{2}-\delta \left(\gamma -\gamma_{0}\right)\right)f_{\gamma}(\gamma)d\gamma\\
&=\int_{0}^{\alpha}f_{\gamma}(\gamma)d\gamma
\!+\!\left(\frac{1}{2}\!+\!\delta\gamma_{0}\right)\int_{\alpha}^{\beta}f_{\gamma}(\gamma) d\gamma\!-\!\delta\int_{\alpha}^{\beta}\gamma f_{\gamma}(\gamma)d\gamma\\
&=F_{\gamma}(\alpha)+\left(\frac{1}{2}+\delta \gamma_{0}\right)
\left(F_{\gamma}(\beta)-F_{\gamma}(\alpha)\right)-\delta \int_{\alpha}^{\beta} \gamma f_{\gamma}(\gamma)d\gamma\\
&\overset{(a)}{=}\left(\frac{1}{2}-\delta \gamma_{0}+\delta \alpha \right) F_{\gamma}(\alpha)
+\left(\frac{1}{2}+\delta \gamma_{0}-\delta \beta \right)F_{\gamma}(\beta)
\notag\\
&~~~~
+\delta\int_{\alpha}^{\beta}F_{\gamma}(\gamma)d\gamma,\label{eq:esilon_approximation_part_e}
\end{align}
\end{subequations}
where step ($a$) is obtained by $\int_{\alpha}^{\beta} \gamma f_{\gamma}(\gamma) d\gamma =\int_{\alpha}^{\beta} \gamma  d F_{\gamma}(\gamma)=\beta F_{\gamma}(\beta)-\alpha F_{\gamma}(\alpha)-\int_{\alpha}^{\beta}  F_{\gamma}(\gamma) d\gamma$.

In our considered truncated CIPC scheme, the non-zero decoding error occurs only when the transmission is performed. As such, for the truncated CIPC scheme, the decoding error probability is conditioned on that the transmit power is not zero (i.e., the channel condition satisfies $\|\mathbf{h}_{\mathrm{u}}\|^{2} \geq \frac{Q}{P_{\mathrm{max}}}$). Therefore, following \eqref{eq:epsilon_def}, the conditional average decoding error probability of the truncated CIPC scheme is given by
\begin{align}\label{eq:epsilon}
&\epsilon\left(Q\middle|\|\mathbf{h}_{\mathrm{u}}\|^{2}\geq\frac{Q}{P_{\mathrm{max}}}\right)
\notag\\
&=\mathbb{E}_{\{\mathbf{h}_{\mathrm{u}}, \mathbf{e}\}}\!\!\left[\!f\left(\!\frac{\sqrt{T}\left(\ln(1+\gamma)\!-\!R\ln2 \right)}{\sqrt{1-\left(1+\gamma\right)^{-2}}}\middle|\|\mathbf{h}_{\mathrm{u}}\|^{2}\geq \frac{Q}{P_{\mathrm{max}}}\!\right)\!\right].
\end{align}

Considering the packet loss caused by both the transmission suspension at the BS and the conditional non-zero decoding error probability at the user, the packet loss probability for our considered truncated CIPC scheme is given by
\begin{align}\label{Define:Packet_Loss_definition}
P_{\epsilon}(Q)=\epsilon\left(Q\middle|\|\mathbf{h}_{\mathrm{u}}\|^{2}\geq \frac{Q}{P_{\mathrm{max}}}\right)p_t(Q)+1-p_t(Q).
\end{align}
We first find that the packet loss probability $P_{\epsilon}(Q)$ is a monotonically increasing function of the transmission rate $R$, since $\epsilon \left(Q\middle|\|\mathbf{h}_{\mathrm{u}}\|^{2}\geq \frac{Q}{P_{\mathrm{max}}}\right)$ monotonically increases when $R$ increases while $p_t(Q)$ is not a function of $R$. Then, we find that $p_t(Q)$ monotonically decreases when $Q$ increases while $\epsilon \left(Q\middle|\|\mathbf{h}_{\mathrm{u}}\|^{2}\geq \frac{Q}{P_{\mathrm{max}}}\right)$ decreases with when $Q$ increases. In other words, there exists an optimal value of $Q$ that minimizes $P_{\epsilon}(Q)$ for the truncated CIPC scheme, which will be tackled in Section~\ref{sec:analysis}. We further find that for the conventional CIPC scheme, we have $p_t(Q) = 1$ and the condition $\|\mathbf{h}_{\mathrm{u}}\|^{2} \geq \frac{Q}{P_{\mathrm{max}}}$ is always satisfied, which leads to the fact that the packet loss probability of the conventional CIPC scheme is the same as the average decoding error probability given in \eqref{eq:epsilon_def}.

\section{Truncated and Conventional Channel Inversion Power Control with Imperfect Channel Reciprocity}\label{sec:analysis}

In this section, we analyze the packet loss probability of the truncated CIPC scheme with imperfect channel reciprocity (i.e., $0 <\phi <1$). Specifically, we determine an upper bound on the receive signal power (i.e., the value of $Q$), which provides a closed interval for the optimal value of $Q$ that minimizes the packet loss probability of the truncated CIPC scheme. Moreover, an easy-to-calculate expression for this packet loss probability $P_{\epsilon}$ is derived in this section. Furthermore, we analyze the conventional CIPC scheme where the BS is not subject to the maximum transmit power constraint, for the sake of performance comparison.

\subsection{Packet Loss Probability of Truncated CIPC Scheme with Imperfect Channel Reciprocity}

When the maximum transmit power constraint is considered, the packet loss probability is affected by the transmission probability and the decoding error probability. We note that, for fixed $R$ and $T$, the packet loss probability $P_{\epsilon}(Q)$ given in \eqref{Define:Packet_Loss_definition} depends on $Q$ heavily. Intuitively, there exists an optimal value of $Q$ that minimizes $P_{\epsilon}(Q)$, since $p_t(Q)$ monotonically decreases when $Q$ increases while $\epsilon \left(Q\middle|\|\mathbf{h}_{\mathrm{u}}\|^{2}\geq \frac{Q}{P_{\mathrm{max}}}\right)$ decreases when $Q$ increases. Therefore, in the following we first derive an approximated but easy-to-calculate expression for $P_{\epsilon}(Q)$ and then use it to determine the optimal value of $Q$ that minimizes $P_{\epsilon}(Q)$.

In order to derive the approximated expression for $P_{\epsilon}(Q)$, we rewrite the SINR given in \eqref{eq:SINRTruncatedCIPC_new} as
\begin{align}\label{eq:SINRTruncatedCIPC}
\gamma=
%\frac{\phi Q}{\left(1-\phi\right)\frac{Q}{{\left\|\mathbf{h}_{\mathrm{u}}\right\|}^2}
%\frac{{\left|\mathbf{e}^{\mathcal{T}}\mathbf{h}_{\mathrm{u}}^{\ast}\right|}^2}
%{{\left\|\mathbf{h}_{\mathrm{u}}\right\|}^2}+\sigma_{w}^{2}}\notag\\
%&=\frac{\phi Q}{Q\left(1-\phi\right)\frac{Y}{X}+\sigma_{w}^{2}}\notag\\
\frac{\phi Q}{Q\left(1-\phi\right)Z+\sigma_{w}^{2}},
\end{align}
where we define $Z=\frac{Y}{X}$ with $X={\left\|\mathbf{h}_{\mathrm{u}}\right\|}^2$ and $Y=\frac{{\left|\mathbf{e}^{\mathcal{T}} \mathbf{h}_{\mathrm{u}}^{\ast} \right|}^2}{{\left\|\mathbf{h}_{\mathrm{u}}\right\|}^2}={\left|\mathbf{e}^{\mathcal{T}} \frac{\mathbf{h}_{\mathrm{u}}^{\ast}}{{\left\|\mathbf{h}_{\mathrm{u}}\right\|}} \right|}^2$. We note that $X$ is independent of $Y$.
%where $\mathbf{h}_{\mathrm{u}}$ and $\mathbf{e}$ are independent and $\mathbf{h}_{\mathrm{u}}\sim\mathcal{CN} \left(0,\mathbf{I}_{N_{\mathrm{t}}}\right)$ i.e., the entries are assumed to be independent and identically distributed (i.i.d.) circularly symmetric complex Gaussian random variables with zero mean and unit variance.
This is due to the fact that $X$ depends on the norm of $\mathbf{h}_{\mathrm{u}}$ but not the phase, while $Y$ depends on the phase of $\mathbf{h}_{\mathrm{u}}$ but not the norm. Since the norm and phase of $\mathbf{h}_{\mathrm{u}}$ are independent, we conclude that $X$ and $Y$ are independent.

We find that in order to calculate the decoding error probability $\epsilon \left(Q\middle|\|\mathbf{h}_{\mathrm{u}}\|^{2}\geq\frac{Q}{P_{\mathrm{max}}}\right)$, we need to derive the conditional cumulative distribution function (CDF) of the SINR. Thus, we next derive this conditional CDF with the maximum transmit power constraint, denoted by $F_{\gamma}\left(\gamma\middle|{\left\|\mathbf{h}_{\mathrm{u}}\right\|}^2\geq\frac{Q}{P_{\mathrm{max}}}\right)$, and present it in the following lemma.

\begin{lemma}\label{Lemma:Distribution_SINR_with_condition}
The conditional CDF of the SINR given in \eqref{eq:SINRTruncatedCIPC} with the maximum transmit power constraint, $F_{\gamma}\left(\gamma\middle|{\left\|\mathbf{h}_{\mathrm{u}}\right\|}^2\geq\frac{Q}{P_{\mathrm{max}}}\right)$, is given by
%%%%%%
\begin{align}\label{eq:CDF_SINR_with_condition}
&F_{\gamma}\left(\gamma\middle|{\left\|\mathbf{h}_{\mathrm{u}}\right\|}^2\geq\frac{Q}{P_{\mathrm{max}}}\right)
%=1-\frac{1}{\Gamma(N_{\mathrm{t}})}
%%\notag\\
%%&~~~~\times
%\left(\gamma_\mathrm{up}\left(N_{\mathrm{t}},\frac{Q}{P_{\mathrm{max}}}\right)
%-\frac{\gamma_\mathrm{up}\left(N_{\mathrm{t}},\frac{Q(1+\xi(\gamma))}
%{P_{\mathrm{max}}}\right)}{\left(1+\xi(\gamma)\right)^{N_{\mathrm{t}}}}\right),
=\frac{\gamma_\mathrm{up}\left(N_{\mathrm{t}},\frac{Q(1+\xi(\gamma))}
{P_{\mathrm{max}}}\right)}{\gamma_\mathrm{up}\left(N_{\mathrm{t}},\frac{Q}{P_{\mathrm{max}}}\right) \left(1+\xi(\gamma)\right)^{N_{\mathrm{t}}}},
\end{align}
where $\xi(\gamma)=\frac{Q\phi-\gamma\sigma_{w}^2}{Q\gamma(1-\phi)}$, $\gamma_\mathrm{lw}(s,x)=\int_{0}^{x} e^{-t} t^{s-1} dt$ is the lower incomplete gamma function \cite[Eq. (8.350.1)]{Gradshteyn2007Book}, $\gamma_\mathrm{up}(s,x)=\int_{x}^{\infty} e^{-t} t^{s-1} dt$ is the upper incomplete gamma function \cite[Eq. (8.350.2)]{Gradshteyn2007Book}, and $\gamma_\mathrm{lw}(s,x)+\gamma_\mathrm{up}(s,x)=\Gamma(s)$ \cite[Eq. (8.356.3)]{Gradshteyn2007Book}.
\begin{IEEEproof}
The detailed proof is presented in Appendix~\ref{Proof:Distribution_SINR_with_condition}.
\end{IEEEproof}
\end{lemma}

We note that a closed-form expression for \eqref{eq:epsilon} is mathematically intractable. Therefore, in the following theorem, we derive an approximated but easy-to-calculate expression for the packet loss probability of the truncated CIPC scheme with the aid of Lemma~\ref{Lemma:Distribution_SINR_with_condition}.
\begin{theorem}\label{Theorem_P_epsilon}
The packet loss probability of the truncated CIPC scheme with imperfect channel reciprocity in URLLC scenarios is approximated as
\begin{align}\label{eq:CloseFormPrEpsilon}
P_{\epsilon}(Q)
&=\Biggl[\left(\frac{1}{2}-\delta \gamma_{0}+\delta\alpha\right) F_{\gamma}\left(\alpha \middle|x \geq \frac{Q}{P_{\mathrm{max}}}\right)
\notag\\
&\hspace{5mm}
+\left(\frac{1}{2}+\delta\gamma_{0}-\delta\beta\right)F_{\gamma}\left(\beta \middle|x\geq \frac{Q}{P_{\mathrm{max}}}\right)\notag\\
&\hspace{5mm}+\delta\int_{\alpha}^{\beta}F_{\gamma}\left(\gamma\middle|x\geq \frac{Q}{P_{\mathrm{max}}}\right)d\gamma\Biggr]
\notag\\
&\hspace{5mm}\times
\left(1-\frac{\gamma_\mathrm{lw}\left(N_{\mathrm{t}},
\frac{Q}{P_{\mathrm{max}}}\right)}{\Gamma(N_{\mathrm{t}})}\right)
+\frac{\gamma_\mathrm{lw}\left(N_{\mathrm{t}},\frac{Q}{P_{\mathrm{max}}}\right)}{\Gamma(N_{\mathrm{t}})},
\end{align}
where $F_{\gamma}\left(\gamma \middle|x \geq \frac{Q}{P_{\mathrm{max}}}\right)$ is given in \eqref{eq:CDF_SINR_with_condition}.
\begin{IEEEproof}
In order to prove Theorem~\ref{Theorem_P_epsilon} and derive the expression for $P_{\epsilon}$, we need to derive the explicit expressions for $p_t(Q)$ and $\epsilon \left(Q\middle|\|\mathbf{h}_{\mathrm{u}}\|^{2} \geq \frac{Q}{P_{\mathrm{max}}}\right)$, which are detailed as follows:

We first tackle the transmission probability $p_t(Q)$. By substituting \eqref{eq:TransmitPowerAP} into \eqref{trans_prob}, we express $p_t(Q)$ as
\begin{align}\label{eq:CloseForm_P_t}
p_t(Q)&=1-\mathrm{Pr}\left\{\|\mathbf{h}_{\mathrm{u}}\|^{2} \leq \frac{Q}{P_{\mathrm{max}}}\right\}
\notag\\
&=1-\frac{\gamma_\mathrm{lw}\left(N_{\mathrm{t}},\frac{Q}{P_{\mathrm{max}}}\right)}{\Gamma(N_{\mathrm{t}})},
\end{align}
%%%%%%%%%
where $f_{X}(x)=\frac{x^{N_\mathrm{t}-1}e^{-x}}{\Gamma(N_\mathrm{t})}$ and $F_{X}(x)=\frac{\gamma_\mathrm{lw}\left(N_{\mathrm{t}},x\right)}{\Gamma(N_{\mathrm{t}})}$ are the probability density function (PDF) and CDF of $\|\mathbf{h}_{\mathrm{u}}\|^{2}$, respectively.

We next write the expression for the decoding error probability in \eqref{eq:epsilon} as
\begin{subequations}
\begin{align}
&\epsilon\left(Q\middle|\|\mathbf{h}_{\mathrm{u}}\|^{2} \geq \frac{Q}{P_{\mathrm{max}}}\right)
\notag\\
&=\mathbb{E}_{\gamma}\left[f\left(\frac{\sqrt{T}\left(\ln(1+\gamma)-R\ln2\right)}
{\sqrt{1-\left(1+\gamma\right)^{-2}}}\middle|x \geq\frac{Q}{P_{\mathrm{max}}}\right)\right]\\
&\overset{(b)}{=}\int_{0}^{\infty} f\left(\frac{\sqrt{T}\left(\ln(1+\gamma)-R\ln2\right)}
{\sqrt{1-\left(1+\gamma\right)^{-2}}}\middle|x \geq \frac{Q}{P_{\mathrm{max}}}\right)
\notag\\
&~~~~~~~~~\times
f_{\gamma}\left(\gamma\middle|x \geq \frac{Q}{P_{\mathrm{max}}}\right)d\gamma\\
&\overset{(c)}{\approx}\int_{0}^{\alpha} f_{\gamma}\left(\gamma \middle|x \geq \frac{Q}{P_{\mathrm{max}}}\right) d\gamma
\notag\\
&~~~~
+\int_{\alpha}^{\beta}\left(\frac{1}{2}-\delta \left(\gamma -\gamma_{0}\right)\right)f_{\gamma}\left(\gamma \middle|x \geq \frac{Q}{P_{\mathrm{max}}}\right) d\gamma
\\\label{eq:Proof-espsilon-condition}
&\overset{(d)}{=}\left(\frac{1}{2}-\delta \gamma_{0}+\delta\alpha\right) F_{\gamma}\left(\alpha \middle|x \geq \frac{Q}{P_{\mathrm{max}}}\right)
\notag\\
&~~~~~~~~~
+\left(\frac{1}{2}+\delta \gamma_{0}-\delta \beta \right)F_{\gamma}\left(\beta \middle| x \geq \frac{Q}{P_{\mathrm{max}}}\right)\notag\\
&~~~~~~~~~+\delta \int_{\alpha}^{\beta}F_{\gamma}\left(\gamma \middle| x \geq \frac{Q}{P_{\mathrm{max}}}\right) d\gamma,
\end{align}
\end{subequations}
where step ($b$) is obtained due to the fact that the transmission condition only has an impact on the distribution, but not the range, of $\gamma$. Steps ($c$) and ($d$) are achieved due to the results given in \eqref{eq:esilon_approximation_part_e}.
Finally, substituting \eqref{eq:CloseForm_P_t} and \eqref{eq:Proof-espsilon-condition} into \eqref{Define:Packet_Loss_definition}, we obtain the desired result in \eqref{eq:CloseFormPrEpsilon}.
\end{IEEEproof}
\end{theorem}

\subsection{Optimization of $Q$ for Truncated CIPC Scheme with Imperfect Channel Reciprocity}

In this subsection, we determine the optimal value of $Q$ to minimize the packet loss probability $P_{\epsilon}(Q)$ of the truncated CIPC scheme for given $T$, $R$, and $P_{\mathrm{max}}$. The  optimization problem at the BS is given by
\begin{align}\label{Objective Function_General}
&\min\limits_{Q}~P_{\epsilon}(Q)
%&~~\textrm{s.t.}
%~~T \geq 100\label{contraints:Blocklength},
\end{align}
%\begin{subequations}\label{Objective Function_General}
%\begin{align}
%&\min\limits_{Q}~P_{\epsilon}(Q)\tag{\ref{Objective Function_General}}
%%&~~\textrm{s.t.}
%%~~T \geq 100\label{contraints:Blocklength},
%\end{align}
%\end{subequations}
%where \eqref{contraints:Blocklength} is the condition for achieving effective approximation for the maximal achievable rate in the finite blocklength regime \cite{Polyanskiy2010}.
%We note that current practical codes have typical blocklength from 100 to 1000 for short-packet control information transmission, e.g., 168 in \cite{Durisi2016TOC}. Hence, \eqref{contraints:Blocklength} can be applied into practical scenarios.

Due to the high complexity involved in the expression for $P_{\epsilon}(Q)$ in \eqref{eq:CloseFormPrEpsilon}, it may not be easy to analytically solve the optimization problem in \eqref{Objective Function_General} by using \eqref{eq:CloseFormPrEpsilon}. To cope with this, we present the following lemma to determine an upper bound on $Q$ for the packet loss probability of the truncated CIPC scheme in the context of URLLC, which ultimately helps to numerically solve the optimization problem.
\begin{lemma}\label{lemma:upper_bound_of_Q}
In the context of URLLC, the value of $Q$ in the truncated CIPC scheme is smaller than a specific value, i.e., $Q<P_{\mathrm{max}}(N_{\mathrm{t}}-1)$ with $N_t > 1$. It is noted that the value of $Q$ is determined by the maximum transmit power and the number of transmit antennas.
\begin{IEEEproof}
We note that the transmission probability $p_t(Q)$ in \eqref{trans_prob} monotonically decreases when $Q$ increases, which is due to the fact that the first-order partial derivative of $p_t(Q)$ w.r.t. $Q$ is given by
\begin{align}\label{fod:P_t}
\frac{\partial \{p_t(Q)\}}{\partial Q}&=-\frac{e^{-\frac{Q}{P_{\mathrm{max}}}} \left(\frac{Q}{P_{\mathrm{max}}}\right)^{N_{\mathrm{t}}-1}}{P_{\mathrm{max}}\Gamma(N_{\mathrm{t}})}<0.
\end{align}
We also note that the second-order partial derivative of $p_t(Q)$ w.r.t. $Q$ is given by
\begin{align}\label{sod:P_t}
\frac{\partial^2 \{p_t(Q)\}}{\partial Q^2}
&=\frac{e^{-\frac{Q}{P_{\mathrm{max}}}} \left(\frac{Q}{P_{\mathrm{max}}}\right)^{N_{\mathrm{t}}+1}
\left(Q-P_{\mathrm{max}}\left(N_{\mathrm{t}}-1\right)\right)}{Q^3\Gamma(N_{\mathrm{t}})}.
\end{align}
As such, we note that the sign of $\frac{\partial^2 \{p_t(Q)\}}{\partial Q^2}$ has three possibilities, which are given by
\begin{align}
\frac{\partial^2 \{p_t(Q)\}}{\partial Q^2} &
\begin{cases}
< 0 ,& \textrm{when}~0 < Q < P_{\mathrm{max}}(N_{\mathrm{t}}-1), \\
= 0 ,& \textrm{when}~Q = P_{\mathrm{max}}(N_{\mathrm{t}}-1), \\
>0 ,& \textrm{when}~Q > P_{\mathrm{max}}(N_{\mathrm{t}}-1).
\end{cases}
\end{align}

It is worth mentioning that $Q$ needs to satisfy the condition of $0<Q<P_{\mathrm{max}}(N_{\mathrm{t}}-1)$. This is due to the fact that when $Q = P_{\mathrm{max}}(N_{\mathrm{t}}-1)$, as per \eqref{eq:CloseForm_P_t} the probability $p_t(Q)$ becomes a function of $N_{\mathrm{t}}$ only, which is given by
\begin{align}
p_t(Q)=p_t(P_{\mathrm{max}}(N_{\mathrm{t}}-1))
=1-\frac{\gamma_\mathrm{lw}\left(N_{\mathrm{t}},N_{\mathrm{t}}-1\right)}
{\Gamma\left(N_{\mathrm{t}}\right)}.
\end{align}
%%%%%%%%%
We note that $p_t(P_{\mathrm{max}}(N_{\mathrm{t}}-1))$ is a monotonically decreasing function of $N_{\mathrm{t}}$ and thus $1 - p_t(P_{\mathrm{max}}(N_{\mathrm{t}}-1))$ increases and approaches a constant value (i.e., $0.5$) when $N_{\mathrm{t}}$ increases. Following \eqref{Define:Packet_Loss_definition}, we have $P_{\epsilon}(Q) > 1 - p_t(P_{\mathrm{max}}(N_{\mathrm{t}}-1))$.
As such, we note that using more transmit antennas is not beneficial to improve reliability when $Q=P_{\mathrm{max}}(N_{\mathrm{t}}-1)$.
We also note that $p_t(Q)=1-\frac{\gamma_\mathrm{lw}\left(\!N_{\mathrm{t}},\frac{Q}{P_{\mathrm{max}}}\!\right)}
{\Gamma(N_{\mathrm{t}})}$ is a monotonically decreasing function of $Q$ due to $\frac{\partial \{p_t(Q)\}}{\partial Q}<0$ proved in \eqref{fod:P_t}. Thus, for $Q > P_{\mathrm{max}}(N_{\mathrm{t}}-1)$, $1-p_t(Q)$ is larger than $0.5$ which violates the requirement of URLLC. As such, for a given $P_{\mathrm{max}}$, we cannot meet the ultra-reliable requirement of URLLC by setting $Q = P_{\mathrm{max}}(N_{\mathrm{t}}-1)$ and we have to decrease $Q$ in order to further increase the value of $p_t(Q)$ due to $\frac{\partial \{p_t(Q)\}}{\partial Q}<0$. Therefore, reducing the value of $Q$ is the only solution to meeting the requirement of URLLC, e.g., satisfying $1 - p_t(Q) \leq 10^{-7}$.
Therefore, in the truncated CIPC scheme, the bound of $Q$ can be determined as $0 < Q < P_{\mathrm{max}}(N_{\mathrm{t}}-1)$, which completes the proof.
\end{IEEEproof}
\end{lemma}

%{With the aid of the upper bound on $Q$ presented in Lemma~\ref{lemma:upper_bound_of_Q}, the optimization of $Q$ in \eqref{Objective Function_General} can be numerically solved by searching the optimal value of $Q$ over the closed interval $[0,  P_{\mathrm{max}}(N_{\mathrm{t}}-1)]$.
%When the reliability requirement of URLLC is given, denoted as $P^{\mathrm{req}}_{\epsilon}$, we can further explore the upper bound on $Q$ by solving $1 - p_t(Q) \leq P^{\mathrm{req}}_{\epsilon}$. It is noted that $1-p_t(Q)=\frac{\gamma_\mathrm{lw}\left(\!N_{\mathrm{t}},\frac{Q}{P_{\mathrm{max}}}\!\right)}
%{\Gamma(N_{\mathrm{t}})}$ is a monotonically increasing function of $Q$ due to $\frac{\partial \{p_t(Q)\}}{\partial Q}<0$ proved in \eqref{fod:P_t}. As such, we can obtain an additional upper bound value on $Q$, denoted by $Q_{\mathrm{up}}$, which is obtained by solving $\frac{\gamma_\mathrm{lw}\left(\!N_{\mathrm{t}},\frac{Q_{\mathrm{up}}}{P_{\mathrm{max}}}\!\right)}
%{\Gamma(N_{\mathrm{t}})}=P^{\mathrm{req}}_{\epsilon}$.
%Then, combining the closed interval in Lemma~\ref{lemma:upper_bound_of_Q}. The new closed interval can be written as $0 < Q < \mathrm{min}\bigl(P_{\mathrm{max}}(N_{\mathrm{t}}-1), Q_{\mathrm{up}}\bigr)$.}
%In addition, we provide a convex set to facilitate determining the optimal value of $Q$ that minimizes the packet loss probability in our proposed truncated CIPC scheme with perfect channel reciprocity, which will be detailed in Section IV.

With the aid of the upper bound on $Q$ presented in Lemma~\ref{lemma:upper_bound_of_Q}, the optimization of $Q$ in \eqref{Objective Function_General} can be numerically solved by searching the optimal value of $Q$ over the closed interval $[0,  P_{\mathrm{max}}(N_{\mathrm{t}}-1)]$.
When the reliability requirement of URLLC is given, denoted as $P^{\mathrm{req}}_{\epsilon}$, we can further explore the upper bound on $Q$ by solving $1 - p_t(Q) \leq P^{\mathrm{req}}_{\epsilon}$. It is noted that $1-p_t(Q)=\frac{\gamma_\mathrm{lw}\left(\!N_{\mathrm{t}},\frac{Q}{P_{\mathrm{max}}}\!\right)}
{\Gamma(N_{\mathrm{t}})}$ is a monotonically increasing function of $Q$ due to $\frac{\partial \{p_t(Q)\}}{\partial Q}<0$ proved in \eqref{fod:P_t}. As such, we can obtain an additional upper bound on $Q$, denoted by $Q_{\mathrm{up}}$, by solving $\frac{\gamma_\mathrm{lw}\left(\!N_{\mathrm{t}},\frac{Q_{\mathrm{up}}}{P_{\mathrm{max}}}\!\right)}
{\Gamma(N_{\mathrm{t}})}=P^{\mathrm{req}}_{\epsilon}$.
Then by combining the closed interval in Lemma~\ref{lemma:upper_bound_of_Q}, we write the new closed interval as $0 < Q < \mathrm{min}\bigl(P_{\mathrm{max}}(N_{\mathrm{t}}-1), Q_{\mathrm{up}}\bigr)$.
In addition, we provide a convex set to determine the optimal value of $Q$ that minimizes the packet loss probability in our proposed truncated CIPC scheme with perfect channel reciprocity, which will be detailed in Section IV.

\subsection{Packet Loss Probability of Conventional CIPC Scheme}

The conventional CIPC scheme is the CIPC scheme without the maximum transmit power constraint, i.e., with $P_{\mathrm{max}} \rightarrow \infty$. This means that the BS can always transmit signals and guarantee $P_{\mathrm{a}}\|\mathbf{h}_{\mathrm{u}}\|^{2}=Q$ in the conventional CIPC scheme. As such, as mentioned before, the transmission probability is one (i.e., $p_t(Q) = 1$) for the conventional CIPC scheme, which leads to the fact that the packet loss probability of the conventional CIPC scheme, i.e., $P^{\infty}_{\epsilon}(Q)$, is same as the decoding error probability, i.e., $\epsilon(\gamma)$ defined in \eqref{eq:epsilon_def}. This enables us to derive an approximated but closed-form expression for the packet loss probability of the conventional CIPC scheme, denoted as $P^{\infty}_{\epsilon}(Q)$, in the following corollary.
\begin{corollary}\label{Corollary_epsilon}
For given finite blocklength $T$ and transmission data rate $R$, the packet loss probability of the conventional CIPC scheme, $P^{\infty}_{\epsilon}(Q)$, is approximated as
\begin{align}\label{eq:epsilon_appro}
P^{\infty}_{\epsilon}(Q)
&=\left(\frac{1}{2}-\delta \gamma_{0}+\delta \alpha \right) \left(\frac{1}{1+\xi(\alpha)}\right)^{N_{\mathrm{t}}}
\notag\\
&~~
+\left(\frac{1}{2}+\delta \gamma_{0}-\delta \beta \right)\left(\frac{1}{1+\xi(\beta)}\right)^{N_{\mathrm{t}}}\notag\\
&~~+\delta M_1^{N_{\mathrm{t}}}\Biggl((-1)^{-N_{\mathrm{t}}} M_2
\biggl(\emph{B}_{-\frac{\alpha}{M_2}}\left(1+N_{\mathrm{t}},1-N_{\mathrm{t}}\right)
\notag\\
&~~~~~~~~~~~~~~
-\emph{B}_{-\frac{\beta}{M_2}}
\left(1+N_{\mathrm{t}},1-N_{\mathrm{t}}\right)\biggr)\Biggr),
\end{align}
where $\xi(x)=\frac{Q \phi-x\sigma_{w}^2}{xQ(1-\phi)}$, $M_1=\frac{Q(1-\phi)}{Q(1-\phi)-\sigma_{w}^2}$, $M_2=\frac{Q \phi}{Q(1-\phi)-\sigma_{w}^2}$, and $\emph{B}_x(a,b)=\int_{0}^{x} t^{a-1} (1-t)^{b-1} dt$ is the incomplete beta function \cite[Eq. (8.391)]{Gradshteyn2007Book}.
\begin{IEEEproof}
The detailed proof is presented in Appendix~\ref{Proof:Corollary_epsilon}.
\end{IEEEproof}
\end{corollary}

We note that the closed-form expression for the packet loss probability of the conventional CIPC scheme offers an upper bound on the performance of the truncated CIPC scheme. This can be used to draw many useful insights with efficient calculations for practical communications scenarios.

\section{Truncated Channel Inversion Power Control with Perfect Channel Reciprocity}\label{sec:performance_comparison}
In this section, we examine the proposed CIPC scheme with perfect channel reciprocity (i.e., $\phi=1$) in the context of URLLC. Specifically, we first derive the packet loss probability for the truncated CIPC scheme with perfect channel reciprocity, which is not a special case of that for the truncated CIPC scheme with imperfect channel reciprocity. We also prove that the packet loss probability of the truncated CIPC scheme with $\phi=1$ is convex w.r.t. $Q$ in a specific set, which significantly facilitates the optimization of $Q$ for the truncated CIPC scheme with perfect channel reciprocity.

\subsection{Packet Loss Probability of Truncated CIPC Scheme with Perfect Channel Reciprocity}

The packet loss probability for the truncated CIPC scheme with perfect channel reciprocity ($\phi=1$) is not a special case of that derived in Theorem \ref{Theorem_P_epsilon} which is for the truncated CIPC scheme with imperfect channel reciprocity ($0<\phi<1$). In addition, the perfect channel reciprocity can exist in some ideal scenarios and aid to obtain an upper bound on the performance of the truncated CIPC scheme in practical scenarios. This motivates us to consider the truncated CIPC scheme with perfect channel reciprocity in this subsection.

Applying perfect channel reciprocity into our proposed truncated CIPC scheme (i.e., $\phi=1$) leads to $\mathbf{h}_{\mathrm{d}}=\mathbf{h}^T_{\mathrm{u}}$. As such, the received signal in \eqref{eq:recei_signal_original} can be rewritten as
\begin{align}\label{eq:recei_signal_perfect_CR}
y_{\phi=1}=\sqrt{P_{\mathrm{a}}}\mathbf{h}_{\mathrm{u}}^{T}\mathbf{x}+w.
\end{align}
Given \eqref{eq:recei_signal_perfect_CR}, the SINR in \eqref{eq:SINR} converts into the SNR given by
\begin{align}\label{eq:SNR}
\gamma_{\phi=1}=\frac{Q}{\sigma_{w}^{2}}.
\end{align}
Based on \eqref{eq:SNR}, we derive the packet loss probability of our proposed truncated CIPC scheme with perfect channel reciprocity in the following lemma.
\begin{lemma}\label{lemma:BLER_perfect_CR}
The packet loss probability of the truncated CIPC scheme with perfect channel reciprocity in URLLC scenarios is derived as
\begin{align}\label{eq:P_e_perfect_CR_closeform}
P^{\phi=1}_{\epsilon}(Q)=& 1-\left(1-\frac{\gamma_\mathrm{lw}\left(N_{\mathrm{t}},\frac{Q}{P_{\mathrm{max}}}\right)}
{\Gamma(N_{\mathrm{t}})}\right)
\notag\\
&\times
\left(1\!-\!f\left(\frac{\sqrt{T}\left(\ln\left(1+\frac{Q}{\sigma_{w}^{2}}\right)-R\ln2 \right)}{\sqrt{1-\left(1+\frac{Q}{\sigma_{w}^{2}}\right)^{-2}}}\right)\right).
\end{align}
\begin{IEEEproof}
In order to prove Lemma~\ref{lemma:BLER_perfect_CR}, we first convert $P_{\epsilon}(Q)$ defined in \eqref{Define:Packet_Loss_definition} into $P^{\phi=1}_{\epsilon}(Q)$, which is given by
\begin{align}\label{eq:P_e_perfect_CR}
P^{\phi=1}_{\epsilon}(Q)=\epsilon\left(Q\right)p_t(Q)+1-p_t(Q).
\end{align}
We note that the transmission probability $p_t(Q)$ is derived in \eqref{eq:CloseForm_P_t}, i.e., $p_t(Q)=1-\frac{\gamma_\mathrm{lw}\left(N_{\mathrm{t}},\frac{Q}{P_{\mathrm{max}}}\right)}
{\Gamma(N_{\mathrm{t}})}$, which is also valid for the perfect channel reciprocity. Then, substituting the SNR, i.e., $\gamma_{\phi=1}$ given by \eqref{eq:SNR}, into \eqref{eq:epsilon_def}, the decoding error probability can be rewritten as
\begin{align}\label{eq:ProofEpsilon}
\epsilon(Q)=f\left(\frac{\sqrt{T}\left(\ln\left(1+\frac{Q}{\sigma_{w}^{2}}\right)-R\ln2 \right)}{\sqrt{1-\left(1+\frac{Q}{\sigma_{w}^{2}}\right)^{-2}}}\right),
\end{align}
where the expectation is eliminated, due to the fact that the SNR in \eqref{eq:SNR} is independent of channel realizations.

Finally, substituting \eqref{eq:CloseForm_P_t} and \eqref{eq:ProofEpsilon}
into \eqref{eq:P_e_perfect_CR}, we obtain the desired result in \eqref{eq:P_e_perfect_CR_closeform}, which completes the proof.
\end{IEEEproof}
\end{lemma}

We note that the packet loss probability $P^{\phi=1}_{\epsilon}(Q)$ is a monotonically increasing function of the transmission rate $R$, since $\epsilon(Q)$ monotonically increases with $R$ while $p_t(Q)$ is not a function of $R$. Meanwhile, $P^{\phi=1}_{\epsilon}(Q)$ monotonically decreases when $P_{\mathrm{max}}$ increases, as $p_t(Q)$ increases when $P_{\mathrm{max}}$ increases, while $\epsilon(Q) <1$ does not depend on $P_{\mathrm{max}}$. In the numerical results, we will examine the required maximum transmit power to achieve URLLC with a certain transmission rate and the maximum allowable packet loss probability.

\subsection{Optimization of $Q$ for Truncated CIPC Scheme with Perfect Channel Reciprocity}

For given $T$, $R$ and $P_{\mathrm{max}}$, the optimization of $Q$ to minimize the packet loss probability in the truncated CIPC scheme with perfect channel reciprocity is given by
\begin{align}\label{Objective Function}
&\min\limits_{Q}~P^{\phi=1}_{\epsilon}(Q)
\end{align}
%\begin{subequations}\label{Objective Function}
%\begin{align}
%&\min\limits_{Q}~P^{\phi=1}_{\epsilon}(Q)\tag{\ref{Objective Function}}\\
%&~~\textrm{s.t.}
%~~T \geq 100.\notag
%\end{align}
%\end{subequations}
Considering the complexity involved in the expression for $P^{\phi=1}_{\epsilon}(Q)$ derived in Lemma~\ref{lemma:BLER_perfect_CR}, it still may not be easy to analytically solve the optimization problem \eqref{Objective Function}. However, we prove that $P^{\phi=1}_{\epsilon}(Q)$ is convex w.r.t. $Q$ in a specific convex set for the optimization problem \eqref{Objective Function}, which is detailed in the following proposition.

\begin{proposition}\label{proposition: optimal_Q}
The packet loss probability $P^{\phi=1}_{\epsilon}(Q)$ of the truncated CIPC is a convex function of $Q$ for $Q_0< Q < P_{\mathrm{max}}(N_{\mathrm{t}}-1)$, where $Q_0=\sigma_{w}^{2}\gamma_a$ and $\gamma_a$ is the solution to
\begin{align}
\frac{\ln(1+\gamma_a)}{(1+\gamma_a})^{2}-1=\frac{1}{3}.
\end{align}
\begin{IEEEproof}
The detailed proof is presented in Appendix~\ref{Proof:proposition_optimal_Q}.
\end{IEEEproof}
\end{proposition}

With the aid of monotonicity and convexity of $P^{\phi=1}_{\epsilon}(Q)$ w.r.t. $Q$ presented in Proposition~\ref{proposition: optimal_Q} and the closed-form expression for $P^{\phi=1}_{\epsilon}(Q)$ derived in Lemma~\ref{lemma:BLER_perfect_CR}, the optimization of $Q$ in \eqref{Objective Function} can be conducted by using some efficient numerical methods, e.g., finding the solution of $Q$ to $\partial \{P^{\phi=1}_{\epsilon}(Q)\}/{\partial Q}=0$ subject to $Q_0< Q < P_{\mathrm{max}}(N_{\mathrm{t}}-1)$ and then comparing the resultant value of $P_{\epsilon}^{\phi=1}(Q)$ with the values of $P_{\epsilon}^{\phi=1}(Q)$ in the region of $0< Q \leq Q_0$.

\section{Numerical Results}\label{sec:numerical results}
In this section, we present numerical results to examine the performance of the proposed truncated CIPC scheme. Based on the numerical results, we draw useful insights into the impact of various system parameters on the performance of the proposed scheme in the considered one-way URLLC scenario. In the following, the value of noise variance is set to $1$, i.e., $\sigma_{w}^{2}=1$.

\begin{figure}[!ht]
\begin{center}
\includegraphics[width=0.98\columnwidth]{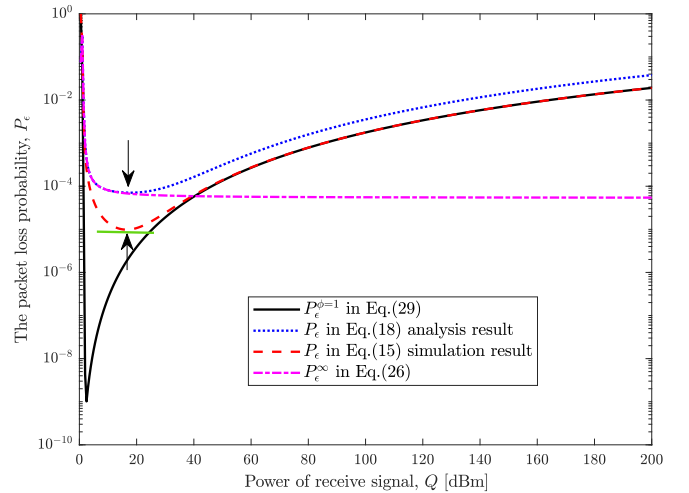}
\caption{The packet loss probability $P_{\epsilon}(Q)$ versus the power of received signals $Q$ in the truncated CIPC scheme with $R=0.8$, $\phi=0.9$, $T=100$, $N_{\mathrm{t}}=4$, and $P_{\mathrm{max}}=23~{\textrm{dBm}}$.}\label{Fig_P_e_vs_Q_accuracy}\vspace{0em}
\end{center}
\end{figure}

Fig.~\ref{Fig_P_e_vs_Q_accuracy} plots the packet loss probabilities, $P_{\epsilon}(Q)$, of the truncated and conventional CIPC schemes versus the power of received signals, $Q$. The simulated and theoretical results are obtained from \eqref{Define:Packet_Loss_definition} and \eqref{eq:CloseFormPrEpsilon}, respectively. The simulated results are obtained by averaging over 10,000 channel realizations. We first observe that there indeed exists an optimal value of $Q$ that minimizes $P_{\epsilon}(Q)$ for the truncated CIPC scheme. As clarified in our analysis, this is mainly due to the maximum transmit power constraint, which results in the fact that both the transmission probability and the decoding error probability decrease when $Q$ increases. We also observe that the trend of theoretical results precisely match the simulated ones in the whole value range of $Q$, which leads to that the optimal value of $Q$ can be precisely searched via our derived easy-to-calculate expression for the packet loss probability of the truncated CIPC scheme given in \eqref{eq:CloseFormPrEpsilon}.

It is noted that in Fig.~\ref{Fig_P_e_vs_Q_accuracy} the curve (with ``legend $P^{\phi=1}_{\epsilon}$ in Eq.~(29)'') is the packet loss probability of the truncated CIPC scheme with perfect channel reciprocity (i.e., $\phi = 1$), while other curves are for imperfect channel reciprocity with $\phi = 0.9$. As expected, we observe that the reliability performance of the truncated CIPC scheme increases when $\phi$ increases and the performance of the truncated CIPC scheme with perfect channel reciprocity serves as an upper bound on this scheme in practical scenarios where the channel reciprocity may not be perfect. Moreover, we observe that the packet loss probability of the conventional CIPC scheme where $P_{\mathrm{max}}\rightarrow\infty$, $P_{\epsilon}^{\infty}$, monotonically decreases when $Q$ increases. This is due to the fact that when the transmit power is unbounded, the transmission always occurs and the transmission probability is one (i.e., $p_t(Q) = 1$). This leads to the fact that the packet loss probability of the conventional CIPC scheme is the same as the average decoding error probability given in \eqref{eq:epsilon_def}. In fact, the decoding error probability is a monotonically decreasing function of $Q$, since the corresponding SINR monotonically increases when $Q$ increases. Therefore, our second observation confirms that our proposed performance metric is more appropriate for one-way URLLC applications in practical wireless scenarios since it considers the transmit power constraint.
Furthermore, we observe that the gap between the minimum packet loss probability of the truncated CIPC scheme (with legend ``$P_{\epsilon}$ in Eq.~(15) simulation results'') that is shown in green curve and the packet loss probability of the conventional CIPC scheme (with legend ``$P^{\infty}_{\epsilon}$ in Eq.~(26)'') is not large and becomes smaller when $Q$ keep increasing.
It is noted that infinite transmit power, i.e., $P_{\mathrm{max}} \rightarrow \infty$ is used in the conventional CIPC scheme, while its achievable minimum packet loss probability is not sensitive to the value of $Q$ and eventually approaches the packet loss probability of the proposed truncated CIPC scheme for given $P_{\mathrm{max}}$ when $Q$ is large enough. In other words, the maximum transmit power plays a critical role in our proposed truncated CIPC scheme for determining the achievable minimum packet loss probability.

%{Here, we needs to verify that when $P_{\mathrm{max}}\rightarrow\infty$, the $P_{\epsilon}^{\infty}(Q)$ of perfect and imperfect channel reciprocities in the truncated CIPC scheme are the same or not, If not. We needs to revises this sentence, e.g., ``For perfect channel reciprocity, the packet loss probability of the conventional CIPC scheme is the same as the average decoding error probability given in \eqref{eq:epsilon_def}. While for imperfect channel reciprocity, the packet loss probability of the conventional CIPC scheme is the same as the \textbf{conditional} average decoding error probability given in \eqref{eq:epsilon}. ''}

%\setcounter{figure}{1}

\begin{figure}[!ht]
\begin{center}
\includegraphics[width=0.98\columnwidth]{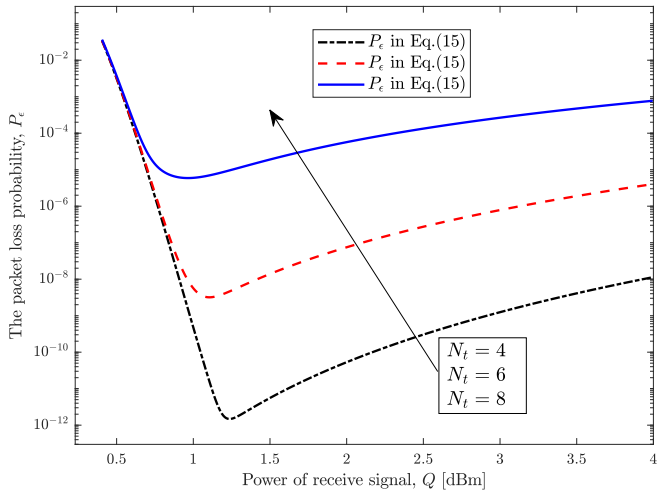}
\caption{The packet loss probability $P_{\epsilon}(Q)$ versus the power of received signal $Q$ in the truncated CIPC scheme for different value of $N_{\mathrm{t}}$ with $R=0.3$, $T=150$, $\phi=0.9$, and $P_{\mathrm{max}}=10~{\textrm{dBm}}$.}\label{Fig_P_e_vs_Q_1}\vspace{0em}
\end{center}
\end{figure}
Fig.~\ref{Fig_P_e_vs_Q_1} plots the packet loss probability $P_{\epsilon}(Q)$ of the truncated CIPC scheme with imperfect channel reciprocity versus the power of received signal $Q$ with different values of $N_{\mathrm{t}}$. First, we observe from this figure the existence of the optimal value of $Q$ that minimizes $P_{\epsilon}(Q)$. Second, we observe that this optimal value is within the interval $(0, P_{\mathrm{max}}(N_{\mathrm{t}}-1))$, which demonstrates the correctness of our Lemma~\ref{lemma:upper_bound_of_Q}.
Third, we observe that the minimum value of $P_{\epsilon}(Q)$ highly depends on the values of $N_{\mathrm{t}}$, i.e., this minimum value decreases significantly when $N_{\mathrm{t}}$ increases. This indicates that the reliability in URLLC can be improved by using more transmit antennas in the truncated CIPC scheme. We note that, without the considered CIPC scheme, increasing the number of transmit antennas may not improve reliability in URLLC. This is due to the fact that the traditional channel estimation overhead increases when there is a higher number of transmit antennas, which limits the reliability performance achieved by using multiple antennas. Fourth, we observe that in the low regime of $Q$, the values of $P_{\epsilon}(Q)$ for different values of $N_{\mathrm{t}}$ are almost the same. Meanwhile, in the high regime of $Q$, the values of $P_{\epsilon}(Q)$ are significantly different for different values of $N_{\mathrm{t}}$.

The main reason lies in the different impact of $N_{\mathrm{t}}$ on $P_{\epsilon}(Q)$ for different $Q$.
Specifically, $P_{\epsilon}(Q)$ is dominated by the conditional decoding error probability $\epsilon\left(Q\middle|\|\mathbf{h}_{\mathrm{u}}\|^{2}\geq \frac{Q}{P_{\mathrm{max}}}\right)$ when $p_t(Q)$ approaches $1$.
It is noted that $\epsilon\left(Q\middle|\|\mathbf{h}_{\mathrm{u}}\|^{2}\geq \frac{Q}{P_{\mathrm{max}}}\right)$ and $p_t(Q)$ are the functions of $N_{\mathrm{t}}$ according to their definitions in \eqref{eq:epsilon} and \eqref{trans_prob}, respectively. To achieve an ultra low decoding error probability, e.g., $10^{-7}$, the value of $p_t(Q)$ needs to be close to $1$ in order to make the packet loss probability $P_{\epsilon}(Q)$ approach the target value. In addition, we note that $p_t(Q)$ and $\epsilon\left(Q\middle|\|\mathbf{h}_{\mathrm{u}}\|^{2}\geq \frac{Q}{P_{\mathrm{max}}}\right)$ are monotonically decreasing functions of $Q$. Therefore, when $Q$ is small, the transmission condition $\|\mathbf{h}_{\mathrm{u}}\|^{2}\geq \frac{Q}{P_{\mathrm{max}}}$ can be easily satisfied, which is not very sensitive to the value of $N_{\mathrm{t}}$.
However, when $Q$ is high, $\|\mathbf{h}_{\mathrm{u}}\|^{2}$ needs to be very large to keep the same transmission condition for given $P_{\mathrm{max}}$. This is due to the fact that the value of $\|\mathbf{h}_{\mathrm{u}}\|^{2}$ significantly depends on the value of $N_{\mathrm{t}}$. In other words, the value of $Q$ heavily affects the impact of $N_{\mathrm{t}}$ on $P_{\epsilon}(Q)$.

\begin{figure}[!ht]
\begin{center}
\includegraphics[width=0.98\columnwidth]{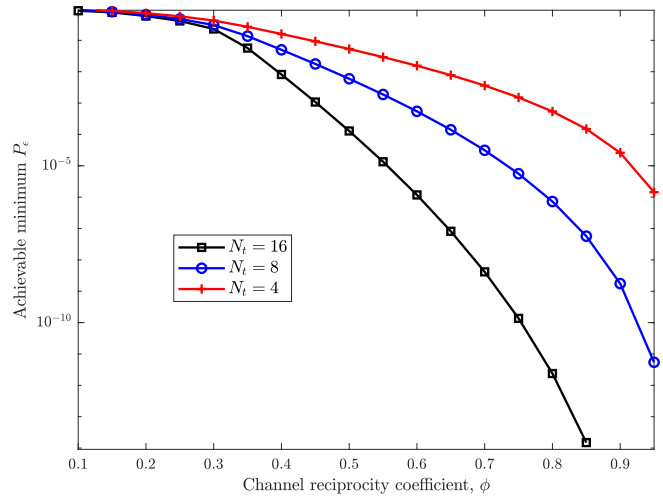}
\caption{The achievable minimum $P_{\epsilon}(Q)$ versus the channel reciprocity coefficient $\phi$ for different number of transmit antennas $N_{\mathrm{t}}$ with $R=0.5$, $T=150$, and $P_{\mathrm{max}}=23~{\textrm{dBm}}$.}\label{Fig_P_e_min_vs_phi}\vspace{0em}
\end{center}
\end{figure}

Fig.~\ref{Fig_P_e_min_vs_phi} plots the achievable minimum $P_{\epsilon}(Q)$ of the truncated CIPC scheme versus the channel reciprocity coefficient $\phi$ for different values of $N_{\mathrm{t}}$.
We first observe that for a large number of transmit antennas, e.g., $N_{\mathrm{t}}=16$, the minimum packet loss probability significantly decreases when $\phi$ increases. This shows the benefits of using multiple antennas at the BS in the considered CIPC schemes for achieving reliability improvement in URLLC scenarios. We also observe that the slope of the minimum packet loss probability with respect to $\phi$ increases when $N_{\mathrm{t}}$ increases, which demonstrates that the benefit of using channel reciprocity to achieve URLLC becomes more profound when the number of transmit antennas increases. Furthermore, in the simulations used to plot this figure, we confirm that the transmission probability $p_t(Q)$ increases and approaches one as the number of transmit antenna $N_{\mathrm{t}}$ increases.

\begin{figure}[!ht]
\begin{center}
\includegraphics[width=0.98\columnwidth]{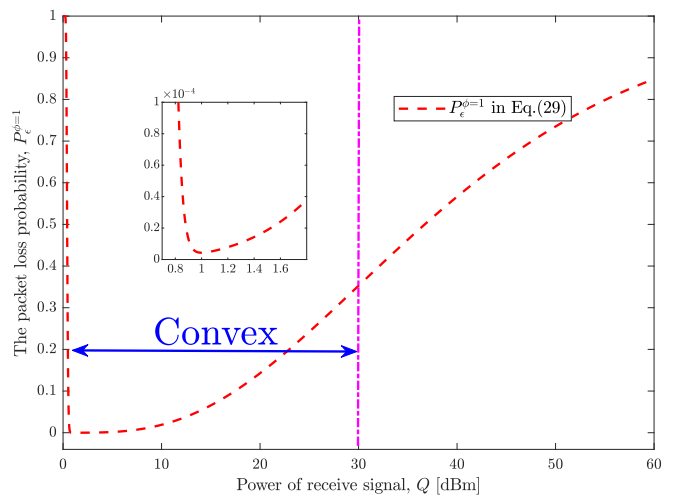}
\caption{The packet loss probability with perfect channel reciprocity, $P^{\phi=1}_{\epsilon}(Q)$, versus the power of receive signal $Q$ in the truncated CIPC scheme with $R=0.5$, $T=150$, $\phi=0.8$, $N_{\mathrm{t}}=4$, and $P_{\mathrm{max}}=10~{\textrm{dBm}}$. }\label{Fig_convexity_P_epsilon}\vspace{0em}
\end{center}
\end{figure}

\begin{figure}[!ht]
\begin{center}
\includegraphics[width=0.98\columnwidth]{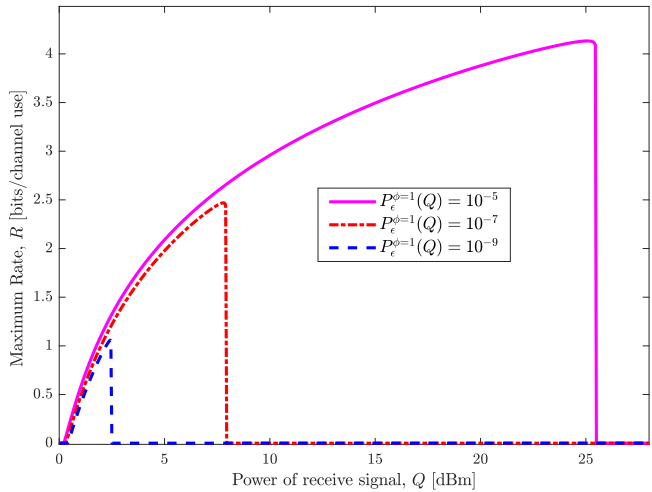}
\caption{The maximum rate $R$ versus the power of received signal $Q$ in the truncated CIPC scheme with perfect channel reciprocity, i.e., $\phi=1$, for different packet loss probabilities $P^{\phi=1}_{\epsilon}(Q)$ with $T=150$, $N_{\mathrm{t}}=4$, and $P_{\mathrm{max}}=23~{\textrm{dBm}}$.}\label{Fig_R_vs_Q}\vspace{0em}
\end{center}
\end{figure}

\begin{figure}[!ht]
\begin{center}
\includegraphics[width=0.98\columnwidth]{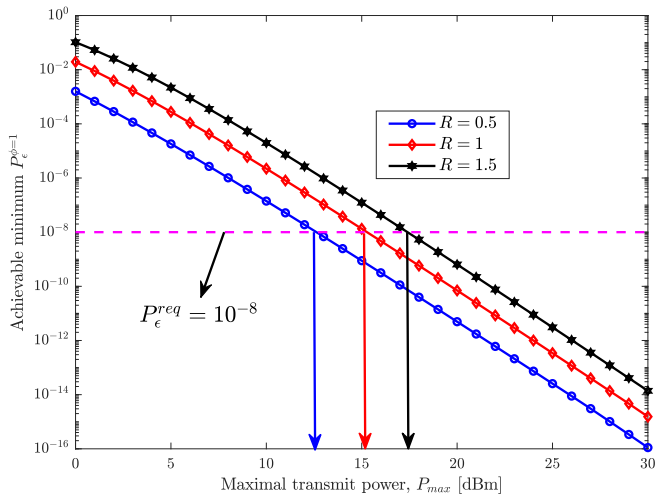}
\caption{The minimum packet loss probability $P^{\phi=1}_{\epsilon}(Q)$ versus the maximum transmit power $P_{\textrm{max}}$ for different values of $R$ with $N_{\mathrm{t}}=5$ and $T=150$.}\label{Fig:P_max}\vspace{0em}
\end{center}
\end{figure}

\begin{figure}[!ht]
\begin{center}
\includegraphics[width=0.98\columnwidth]{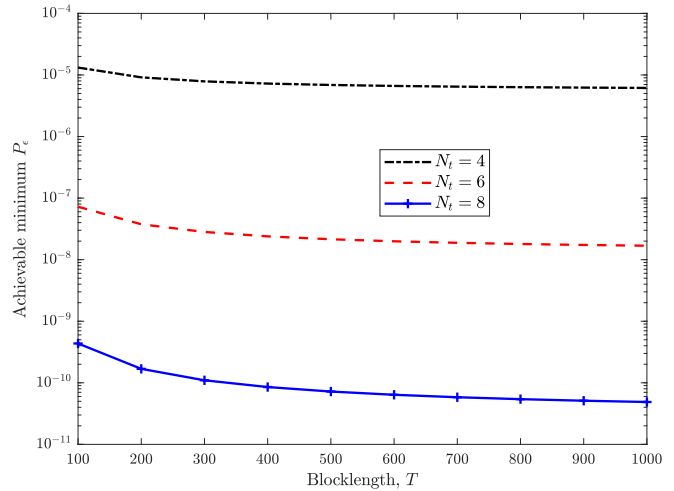}
\caption{The minimum packet loss probability $P_{\epsilon}(Q)$ versus the blocklength $T$ in the truncated CIPC scheme with $R=0.8$, $\phi=0.9$, and $P_{\mathrm{max}}=23~{\textrm{dBm}}$ for different $N_{\mathrm{t}}$.}\label{Fig_P_e_min_vs_T}\vspace{0em}
\end{center}
\end{figure}

Fig.~\ref{Fig_convexity_P_epsilon} plots the packet loss probability of the truncated CIPC scheme with perfect channel reciprocity $P^{\phi=1}_{\epsilon}(Q)$ versus the power of received signal $Q$. As shown in Fig.~\ref{Fig_convexity_P_epsilon}, we observe that the optimal value of $Q$ that minimizes $P^{\phi=1}_{\epsilon}(Q)$ is within the interval $(Q_0, P_{\mathrm{max}}(N_{\mathrm{t}}-1))$, where $P_{\mathrm{max}}(N_{\mathrm{t}}-1)=30$ in this figure. More specifically, the figure shows that $P^{\phi=1}_{\epsilon}(Q)$ first decreases and then increases when $Q$ increases. The result demonstrates the correctness of Proposition~\ref{proposition: optimal_Q}, where $P^{\phi=1}_{\epsilon}(Q)$ of the truncated CIPC is a convex function of $Q$ for $Q_0< Q < P_{\mathrm{max}}(N_{\mathrm{t}}-1)$.

Fig.~\ref{Fig_R_vs_Q} plots the maximum rate $R$ versus the power of received signal $Q$ in the truncated CIPC scheme with perfect channel reciprocity, i.e., $\phi=1$, for different reliability requirements, where $P^{\phi=1}_{\epsilon}(Q)=10^{-5}, 10^{-7}~\text{and}~ 10^{-9}$.
As shown in Fig.~\ref{Fig_R_vs_Q}, the maximum achievable rate decreases when the reliability requirement becomes more stringent. Moreover, we find that the maximum value of $R$ that guarantees the target $P^{\phi=1}_{\epsilon}(Q)$ first increases nonlinearly with $Q$ and then immediately reduces to zero as $Q$ becomes larger. This is due to the fact that $P^{\phi=1}_{\epsilon}(Q)$ first decreases and then increases when $Q$ increases. In other words, when $Q$ is larger than the threshold that satisfying the reliability requirement, no achievable rate can be supported by current system parameters.

Fig.~\ref{Fig:P_max} plots the minimum packet loss probability of the truncated CIPC scheme with perfect channel reciprocity versus the maximum transmit power $P_{\textrm{max}}$ for different values of the transmission rate $R$. As expected, we first observe that the minimum packet loss probability monotonically decreases when $P_{\textrm{max}}$ increases, since increasing $P_{\textrm{max}}$ enables the BS to send signals under more channel conditions and thus increases the transmission probability $p_t(Q)$. This demonstrates that the maximum transmit power plays a critical role in the truncated CIPC scheme. This figure also demonstrates that to guarantee a certain reliability, the required value of $P_{\textrm{max}}$ increases when the transmission rate $R$ increases, which can be explicitly determined by our examinations. This reveals one specific contribution of this work, i.e., determining system parameters (e.g., $P_{\textrm{max}}$) for given requirements on the considered URLLC scenario. Furthermore, we observe that the minimum packet loss probability increases when $R$ increases, which demonstrates the tradeoff between the transmission rate $R$ and reliability in URLLC.

Fig. \ref{Fig_P_e_min_vs_T} plots the achievable minimum $P_{\epsilon}(Q)$ of the truncated CIPC scheme versus the blocklength $T$ for different values of $N_{\mathrm{t}}$. As expected, we observe that the minimum packet loss probability significantly decreases when $N_{\mathrm{t}}$ increases, which shows the benefits of using multiple antennas at the BS for reliability improvement in URLLC scenarios. In addition, we find that the minimum packet loss probability decreases and tends to be saturated when $T$ becomes large (i.e., $T=1000$). This demonstrates that for given system parameters, the reliability improvement caused by increasing the blocklength becomes negligible when $T$ becomes large.

\section{Conclusions}\label{sec:conclusions}

This work proposed to use the CIPC schemes to achieve one-way URLLC with on-hand performance evaluation. Specifically, we first derived expressions for the packet loss probabilities of the truncated and traditional CIPC schemes with imperfect channel reciprocity. Using these expressions, we determined a closed interval for the optimal value of the received signal power $Q$, which significantly facilitates the optimal design of the CIPC schemes. Then, we analyzed the performance of the truncated CIPC scheme with perfect channel reciprocity, which provides an upper bound on the performance of truncated CIPC in practical scenarios. Based on this analysis, we proved that the optimal $Q$ lies in a convex set in the case with perfect channel reciprocity. Our examination explicitly determined the trade-off among reliability, latency, and required communication resources (e.g., transmit antennas and transmit power), which provides novel design guidelines into achieving one-way URLLC with the CIPC schemes.

\appendices

\section{Proof of Lemma~\ref{Lemma:Distribution_SINR_with_condition}}\label{Proof:Distribution_SINR_with_condition}

The conditional CDF of the SINR with the maximum transmit power constraint is rewritten as
\begin{align}\label{eq:CDF_SINR_1_condition}
&F_{\gamma}\left(\gamma\middle|\|\mathbf{h}_{\mathrm{u}}\|^{2}\geq\frac{Q}{P_{\mathrm{max}}}\right)
\notag\\
&\hspace{10mm}=\text{Pr}\left\{\frac{Q \phi}{Q(1-\phi)\frac{Y}{X}+\sigma_{w}^2}\leq\gamma
\middle|X\geq\frac{Q}{P_{\mathrm{max}}}\right\}
\notag\\
&\hspace{10mm}=1-\text{Pr}\left\{Y\leq X\xi(\gamma)\middle|X\geq\frac{Q}{P_{\mathrm{max}}}\right\},
\end{align}
where $\xi(\gamma)$ is defined below \eqref{eq:CDF_SINR_with_condition}.
Then, we write $\text{Pr}\left\{Y\leq X\xi(\gamma)\middle|X\geq \frac{Q}{P_{\mathrm{max}}}\right\}$ as
\begin{align}\label{eq:CDF_SINR_2_condition}
&\text{Pr}\left\{Y\leq X\xi(\gamma)\middle|X\geq\frac{Q}{P_{\mathrm{max}}}\right\}
%&\hspace{10mm}=\int_{\frac{Q}{P_{\mathrm{max}}}}^{\infty}\text{Pr}\left\{Y\leq x\xi(\gamma) \right\}f_{X}(x)dx
\notag\\
&\hspace{10mm}=\frac{\text{Pr}\left\{Y\leq X\xi(\gamma), X\geq\frac{Q}{P_{\mathrm{max}}}\right\}}{\text{Pr}\left\{X\geq\frac{Q}{P_{\mathrm{max}}}\right\}}
%\notag\\
%&=\frac{\int_{\frac{Q}{P_{\mathrm{max}}}}^{\infty}F_{Y}\bigl(x \xi(\gamma)\bigr)f_{X}(x)dx}
%{\text{Pr}\left\{X\geq\frac{Q}{P_{\mathrm{max}}}\right\}}
\notag\\
&\hspace{10mm}=\frac{\int_{\frac{Q}{P_{\mathrm{max}}}}^{\infty}\left(1-e^{-\left(x \xi(\gamma)\right)}\right)\frac{x^{N_{\mathrm{t}}-1}}{\Gamma(N_{\mathrm{t}})}e^{-x} dx}
{\left(\frac{\gamma_\mathrm{up}\left(N_{\mathrm{t}},\frac{Q}{P_{\mathrm{max}}}\right)}{\Gamma(N_{\mathrm{t}})}\right)}
\notag\\
&\hspace{10mm}=\frac{\frac{1}{\Gamma(N_{\mathrm{t}})}
\left(\gamma_\mathrm{up}\left(N_{\mathrm{t}},\frac{Q}{P_{\mathrm{max}}}\right)
\!-\!\frac{\gamma_\mathrm{up}\left(N_{\mathrm{t}},\frac{Q(1+\xi(\gamma))}
{P_{\mathrm{max}}}\right)}{\left(1+\xi(\gamma)\right)^{N_{\mathrm{t}}}}\right)}
{\left(\frac{\Gamma(N_{\mathrm{t}})}{\gamma_\mathrm{up}\left(N_{\mathrm{t}},\frac{Q}{P_{\mathrm{max}}}\right)}\right)}
\notag\\
&\hspace{10mm}=1-\frac{\gamma_\mathrm{up}\left(N_{\mathrm{t}},\frac{Q(1+\xi(\gamma))}
{P_{\mathrm{max}}}\right)}{\gamma_\mathrm{up}\left(N_{\mathrm{t}},\frac{Q}{P_{\mathrm{max}}}\right) \left(1+\xi(\gamma)\right)^{N_{\mathrm{t}}}},
\end{align}
%%%%%%%%%
where $f_{\mathrm{X}}(x)=\frac{x^{N_{\mathrm{t}}-1}}{\Gamma(N_{\mathrm{t}})} e^{-x}$, $F_{\mathrm{Y}}(y)=1-e^{-y}$, and $\gamma_\mathrm{up}(s,x)$ is defined below \eqref{eq:CDF_SINR_with_condition}. By substituting \eqref{eq:CDF_SINR_2_condition} into \eqref{eq:CDF_SINR_1_condition}, we obtain the result in \eqref{eq:CDF_SINR_with_condition}, which completes the proof.

\section{Proof of Corollary~\ref{Corollary_epsilon}}\label{Proof:Corollary_epsilon}

Following Theorem~\ref{Theorem_P_epsilon} and considering $P_{\mathrm{max}} \rightarrow \infty$, the packet loss probability of the conventional CIPC scheme, $P^{\infty}_{\epsilon}(Q)$, is given by
\begin{align}\label{eq:P_epsilon_P_max_inf}
P^{\infty}_{\epsilon}(Q)
&=\Biggl[\left(\frac{1}{2}-\delta \gamma_{0}+\delta\alpha\right)F_{\gamma}\left(\alpha \middle|x \geq 0\right)
\notag\\
&\hspace{5mm}+\left(\frac{1}{2}+\delta \gamma_{0}-\delta \beta \right)F_{\gamma}\left(\beta\middle|x\geq 0\right)\notag\\
&\hspace{5mm}+\delta\int_{\alpha}^{\beta}F_{\gamma}\left(\gamma \middle|x \geq 0\right) d\gamma\Biggr]
\notag\\
&\hspace{5mm} \times \Biggl(1-\frac{\gamma_\mathrm{lw}\left(N_{\mathrm{t}},0\right)}{\Gamma(N_{\mathrm{t}})}\Biggr)
+\frac{\gamma_\mathrm{lw}\left(N_{\mathrm{t}},0\right)}{\Gamma(N_{\mathrm{t}})}\notag\\
&=\delta\int_{\alpha}^{\beta}F_{\gamma}\left(\gamma\middle|x\geq0\right)d\gamma
\notag\\
&\hspace{5mm}+\left(\frac{1}{2}-\delta\gamma_{0}+\delta \alpha \right)
 F_{\gamma}\left(\alpha\middle|x\geq 0\right)\notag\\
&\hspace{5mm}+\left(\frac{1}{2}+\delta\gamma_{0}-\delta \beta\right)\!F_{\gamma}\left(\beta\middle|x \geq 0\right),
\end{align}
where $\frac{\gamma_\mathrm{lw}\left(N_{\mathrm{t}},\frac{Q}{P_{\mathrm{max}}}\right)}{\Gamma(N_{\mathrm{t}})}
=\frac{\gamma_\mathrm{lw}\left(N_{\mathrm{t}},0\right)}{\Gamma(N_{\mathrm{t}})}=0$ as $P_{\mathrm{max}}\rightarrow \infty$.

Then, we need to derive the CDF of the SINR to solve the integration in \eqref{eq:P_epsilon_P_max_inf}. According to Lemma~\ref{Lemma:Distribution_SINR_with_condition}, we obtain $F_{\gamma}\left(\gamma \middle|x \geq 0\right)$ by substituting $\frac{Q}{P_{\mathrm{max}}}=0$ into \eqref{eq:CDF_SINR_with_condition}, which leads to
\begin{align}\label{eq:CDF_SINR_cond_P_max_inf}
F_{\gamma}\left(\gamma\middle|x \geq 0\right)
%&=1-\frac{1}{\Gamma(N_{\mathrm{t}})}\left(\gamma_\mathrm{up}\left(N_{\mathrm{t}},0\right)
%-\frac{\gamma_\mathrm{up}\left(N_{\mathrm{t}},0\right)}
%{\left(1+\xi(\gamma)\right)^{N_{\mathrm{t}}}}\right)\notag\\
&=1-\frac{1}{\Gamma(N_{\mathrm{t}})}\left(\Gamma(N_{\mathrm{t}})
-\frac{\Gamma(N_{\mathrm{t}})}{\left(1+\xi(\gamma)\right)^{N_{\mathrm{t}}}}\right)
\notag\\
&=\left(\frac{1}{1+\xi(\gamma)}\right)^{N_{\mathrm{t}}},
\end{align}
where $\gamma_\mathrm{up}\left(N_{\mathrm{t}},0\right)=\Gamma(N_{\mathrm{t}})$.
As such, $\int_{\alpha}^{\beta} F_{\gamma}\left(\gamma \middle|x \geq 0\right) d\gamma$ can be obtained as
\begin{align}\label{eq:interation_2}
\int_{\alpha}^{\beta}F_{\gamma}&\left(\gamma \middle|x \geq 0\right)d\gamma
=\int_{\alpha}^{\beta}\left(\frac{1}{1+\xi(\gamma)}\right)^{N_{\mathrm{t}}}d\gamma
\notag\\
&\overset{(e)}{=}\int_{\alpha}^{\beta} \left(\frac{M_1 \gamma}{\gamma+M_2}\right)^{N_{\mathrm{t}}}d\gamma
\notag\\
&=\int_{\alpha}^{\beta} M_1^{N_{\mathrm{t}}}\left(\frac{ \gamma}{\gamma+M_2}\right)^{N_{\mathrm{t}}}d\gamma
\notag\\
&=M_1^{N_{\mathrm{t}}}\Biggl((-1)^{-N_{\mathrm{t}}} M_2
\biggl(\emph{B}_{-\frac{\alpha}{M_2}}\left(1+N_{\mathrm{t}},1-N_{\mathrm{t}}\right)
\notag\\
&\hspace{15mm}-\emph{B}_{-\frac{\beta}{M_2}}\left(1+N_{\mathrm{t}},1-N_{\mathrm{t}}\right)
\biggr)\Biggr),
\end{align}
%%%%%%%%%
where step ($e$) is achieved by using $\frac{1}{1+\xi(\gamma)}=\frac{M_1 \gamma}{\gamma+M_2}$, while $\xi(x)$, $M_1$, $M_2$, and $B_x(a,b)$ are defined below \eqref{eq:epsilon_appro}. Then, we obtain \eqref{eq:epsilon_appro} by substituting \eqref{eq:CDF_SINR_cond_P_max_inf} and \eqref{eq:interation_2} into \eqref{eq:P_epsilon_P_max_inf}, which completes the proof.

\section{Proof of Proposition~\ref{proposition: optimal_Q}}\label{Proof:proposition_optimal_Q}

In order to find the optimal $Q$ to minimize the packet loss probability, $P^{\phi=1}_{\epsilon}(Q)$, we need to find the monotonicity and convexity of $P^{\phi=1}_{\epsilon}(Q)$ w.r.t. $Q$. To this end, we first derive the first-order partial derivative of $P^{\phi=1}_{\epsilon}(Q)$ w.r.t. $Q$, which is given by
\begin{align}\label{fod:P_epsilon}
\frac{\partial \{P^{\phi=1}_{\epsilon}(Q)\} }{\partial Q}
&=\frac{\partial \{p_t(Q)\}}{\partial Q}\biggl(\epsilon(Q)-1\biggr)+ p_t(Q) \frac{\partial \{\epsilon(Q)\}}{\partial Q}.
\end{align}
Then, the second-order partial derivative of $P^{\phi=1}_{\epsilon}(Q)$ w.r.t. $Q$ is obtained as
\begin{align}\label{sod:P_epsilon}
&\frac{\partial^2 \{P^{\phi=1}_{\epsilon}(Q)\} }{\partial Q^2}
=\frac{\partial^2 \{p_t(Q)\}}{\partial Q^2}
\biggl(\epsilon(Q)-1\biggr)
\notag\\
&\hspace{10mm}
+ 2 \frac{\partial \{p_t(Q)\}}{\partial Q}
\frac{\partial \{\epsilon(Q)\}}{\partial Q}
+p_t(Q) \frac{\partial^2 \{\epsilon(Q)\}}{\partial Q^2}.
\end{align}

In order to determine the sign of $\frac{\partial^2 \{P^{\phi=1}_{\epsilon}(Q)\}}{\partial Q^2}$, we first need to tackle $\frac{\partial \{p_t(Q)\}}{\partial Q}$ and $\frac{\partial^2 \{p_t(Q)\}}{\partial Q^2}$.
Recall that $\frac{\partial \{p_t(Q)\}}{\partial Q}<0$ and $\frac{\partial^2 \{p_t(Q)\}}{\partial Q^2}$ have been derived in \eqref{fod:P_t} and \eqref{sod:P_t}, respectively. As such, we find that $p_t(Q)$ is a monotonically decreasing function of $Q$ due to $\frac{\partial \{p_t(Q)\}}{\partial Q}<0$. Besides, with the aid of the proof in Lemma~\ref{lemma:upper_bound_of_Q}, we can obtain that $\frac{\partial \{p_t(Q)\}}{\partial Q}<0$ and $\frac{\partial^2 \{p_t(Q)\}}{\partial Q^2} < 0$ for $0 < Q < P_{\mathrm{max}}(N_{\mathrm{t}}-1)$.
Then, we find the signs of $\frac{\partial \{\epsilon(Q)\}}{\partial Q}$ and $\frac{\partial^2 \{\epsilon(Q)\}}{\partial Q^2}$ as follows:

We first find the sign of $\frac{\partial \epsilon(Q)}{\partial Q}$. To this end, we rewrite the decoding error probability in \eqref{eq:epsilon_def} as $\epsilon(\gamma_{\phi=1})=f(A(\gamma_{\phi=1}))$,
where $A\left(\gamma_{\phi=1}\right)=\frac{\sqrt{T}\left(\ln(1+\gamma_{\phi=1})-R\ln2 \right)}{\sqrt{1-\left(1+\gamma_{\phi=1}\right)^{-2}}}$.
As such, the first-order partial derivative of $\epsilon(Q)$ w.r.t. $Q$ is derived as
\begin{align}\label{fod:Epsilon}
\frac{\partial\{\epsilon(Q)\}}{\partial Q}
&=\frac{\partial\{f(A(\gamma_{\phi=1}))\}}{\partial Q}
\notag\\
&=\frac{\partial\left\{f(A(\gamma_{\phi=1}))\right\} }{\partial \{A(\gamma_{\phi=1})\}}
\frac{\partial\{A(\gamma_{\phi=1})\}}{\partial \gamma_{\phi=1}}
\frac{\partial \gamma_{\phi=1}}{\partial Q}.
\end{align}
We note that the sign of $\frac{\partial \{\epsilon(Q)\}}{\partial Q}$ is determined by the signs of $\frac{\partial \left\{ f(A(\gamma_{\phi=1}))\right\} }{\partial \{A(\gamma_{\phi=1})\}}$, $\frac{\partial \{A(\gamma_{\phi=1})\} }{\partial \gamma_{\phi=1}}$, and $\frac{\partial \gamma_{\phi=1} }{\partial Q}$. As such, we can conclude that $\epsilon(Q)$ is a decreasing function of $Q$ if we can prove $\frac{\partial \left\{ f(A(\gamma_{\phi=1}))\right\} }{\partial \{A(\gamma_{\phi=1})\}} \frac{\partial \{A(\gamma_{\phi=1})\} }{\partial \gamma_{\phi=1}}\frac{\partial \gamma_{\phi=1} }{\partial Q}<0$. In order to prove this, we present the following three results given by
\begin{align}
\frac{\partial \left\{ f(A(\gamma_{\phi=1}))\right\} }{\partial \{A(\gamma_{\phi=1})\}}
&=-\frac{1}{\sqrt{2\pi}}\exp\left(-\frac{A^{2}(\gamma_{\phi=1})}{2}\right)<0, \label{fod:f_A_gamma_wrt_A_gamma}\\
\frac{\partial \{A(\gamma_{\phi=1})\}}{\partial \gamma_{\phi=1}}
&=\frac{\sqrt{T}\left(1-\frac{\ln(1+\gamma_{\phi=1})-R \ln2}{(1+\gamma_{\phi=1})^{2}-1}\right)}
{\sqrt{(1+\gamma_{\phi=1})^{2}-1}}>0,\label{fod:A_gamma_wrt_Q}\\
\frac{\partial \gamma_{\phi=1}}{\partial Q}
&=\frac{1}{\sigma_{w}^{2}}>0.\label{fod:gamma_wrt_Q}
\end{align}
%%%%%%%%%
We note that \eqref{fod:f_A_gamma_wrt_A_gamma} is obtained due to the properties of the $Q$-function and \eqref{fod:A_gamma_wrt_Q} is obtained due to the proof in \cite[Appendix A]{Li2018ICCW}. According to \eqref{fod:f_A_gamma_wrt_A_gamma}, \eqref{fod:A_gamma_wrt_Q}, and \eqref{fod:gamma_wrt_Q}, we obtain $\frac{\partial \{f(A(\gamma_{\phi=1}))\}}{\partial Q}
 = {\frac{\partial \left\{ f(A(\gamma_{\phi=1}))\right\} }{\partial \{A(\gamma_{\phi=1})\}}}
{\frac{\partial \{A(\gamma_{\phi=1})\} }{\partial \gamma_{\phi=1}}}
{\frac{\partial \gamma_{\phi=1} }{\partial Q}}<0$,  which results in $\frac{\partial \{\epsilon(Q)\}}{\partial Q}<0$.
We then find the sign of $\frac{\partial^2 \{\epsilon(Q)\}}{\partial Q^2}$. To this end, we express $ \frac{\partial^2 \{\epsilon(Q)\}}{\partial Q^2}$ as
\begin{align}\label{sod:Epsilon}
\frac{\partial^2 \{\epsilon(Q)\}}{\partial Q^2}
&=\frac{\partial^2 \{f(A(\gamma_{\phi=1}))\}}{\partial Q^2}.
\end{align}
In order to obtain the sign of $\frac{\partial^2 \left\{ f(A(\gamma_{\phi=1}))\right\}}{\partial Q^2}$, we rewrite $\frac{\partial^2 \left\{ f(A(\gamma_{\phi=1}))\right\}}{\partial Q^2}$ as
\begin{align}
&\frac{\partial^2 \left\{ f(A(\gamma_{\phi=1}))\right\}}{\partial Q^2}
\notag\\
%&=\frac{\partial^2 \left\{ f(A(\gamma_{\phi=1}))\right\} }{\partial \{A(\gamma_{\phi=1})\}\partial Q} \frac{\partial \{A(\gamma_{\phi=1})\} }{\partial \gamma_{\phi=1}}  \frac{\partial \gamma_{\phi=1} }{\partial Q}\notag\\
%&~~~+\frac{\partial \left\{ f(A(\gamma_{\phi=1}))\right\} }{\partial \{A(\gamma_{\phi=1})\}} \frac{\partial^2 \{A(\gamma_{\phi=1})\} }{\partial \gamma_{\phi=1} \partial Q}  \frac{\partial \gamma_{\phi=1} }{\partial Q}\notag\\
%&~~~+\frac{\partial \left\{ f(A(\gamma_{\phi=1}))\right\} }{\partial \{A(\gamma_{\phi=1})\}} \frac{\partial \{A(\gamma_{\phi=1})\} }{\partial \gamma_{\phi=1}}  \frac{\partial^2 \gamma_{\phi=1} }{\partial Q^2}\notag\\
&~~~~~=\frac{\partial^2 \left\{ f(A(\gamma_{\phi=1}))\right\} }{\partial \{A^2(\gamma_{\phi=1})\}}
\left\{\frac{\partial \{A(\gamma_{\phi=1})\}}{\partial \gamma_{\phi=1}}\right\}^{2}
\left\{\frac{\partial \gamma_{\phi=1}}{\partial Q}\right\}^{2}\notag\\
&~~~~~~~~~~+\frac{\partial \left\{ f(A(\gamma_{\phi=1}))\right\} }{\partial \{A(\gamma_{\phi=1})\}}
\frac{\partial^2 \{A(\gamma_{\phi=1})\} }{\partial \gamma_{\phi=1}^2}
\left\{\frac{\partial \gamma_{\phi=1}}{\partial Q}\right\}^{2}\notag\\
&~~~~~~~~~~+\frac{\partial \left\{ f(A(\gamma_{\phi=1}))\right\} }{\partial \{A(\gamma_{\phi=1})\}}\frac{\partial\{A(\gamma_{\phi=1})\} }{\partial \gamma_{\phi=1}}
\frac{\partial^2 \gamma_{\phi=1} }{\partial Q^2},
\end{align}
where $\frac{\partial^2 \gamma_{\phi=1} }{\partial Q^2}=0$. Based on the properties of the $Q$-function, we express $\frac{\partial^2 \left\{f(A(\gamma_{\phi=1}))\right\}}{\partial\{A^2(\gamma_{\phi=1})\}}$ as
\begin{align}\label{sod:f_A_gamma_wrt_A_gamma}
\frac{\partial^2 \left\{f(A(\gamma_{\phi=1}))\right\}}{\partial\{A^2(\gamma_{\phi=1})\}}
&\!=\!\frac{A(\gamma_{\phi=1})}{\sqrt{2\pi}}\exp\left(\!-\frac{A^{2}(\gamma_{\phi=1})}{2}\!\right)>0.
\end{align}
For now, we have $\frac{\partial^2 \left\{f(A(\gamma_{\phi=1}))\right\}}{\partial\{A^2(\gamma_{\phi=1})\}}>0$ in \eqref{sod:f_A_gamma_wrt_A_gamma}, $\frac{\partial \left\{ f(A(\gamma_{\phi=1}))\right\} }{\partial \{A(\gamma_{\phi=1})\}}<0$ in \eqref{fod:f_A_gamma_wrt_A_gamma}, $\frac{\partial \{A(\gamma_{\phi=1})\} }{\partial \gamma_{\phi=1}}>0$ in \eqref{fod:A_gamma_wrt_Q}, $\frac{\partial \gamma_{\phi=1} }{\partial Q}>0$ in \eqref{fod:gamma_wrt_Q}, and $\frac{\partial^2 \gamma_{\phi=1}}{\partial Q^2}=0$. Thus, determining the sign of $\frac{\partial^2 \left\{ f(A(\gamma_{\phi=1}))\right\}}{\partial Q^2}$ is equivalent to identifying the sign of $\frac{\partial^2 \{A(\gamma_{\phi=1})\}}{\partial \gamma_{\phi=1}^2}$. As per the proof given in \cite[Appendix A]{Li2019GlobeCOMW}, we prove that $\frac{\partial^2  \{A(\gamma_{\phi=1})\} }{\partial \gamma_{\phi=1}^2}<0$ for $\gamma_{\phi=1}>\gamma_a$, where $\gamma_a$ is the solution to
\begin{align}
\frac{\ln(1+\gamma_a)}{(1+\gamma_a)^{2}-1}=\frac{1}{3}.
\end{align}
Following \eqref{eq:SNR}, $\gamma_{\phi=1}>\gamma_a$ leads to
\begin{align}
\gamma_{\phi=1}>\gamma_a\Longrightarrow
\frac{Q}{\sigma_{w}^{2}} >\gamma_a\Longrightarrow Q>Q_0,
\end{align}
where $Q_0=\sigma_{w}^{2} \gamma_a$
%where $Q_0$ is given by
%\begin{align}\label{eq:Q_0}
%Q_0=\sigma_{w}^{2} \gamma_a.
%\end{align}
As such, for $Q>Q_0$, we have $\frac{\partial^2 \left\{ f(A(\gamma_{\phi=1}))\right\}}{\partial Q^2}>0$, due to
\begin{align}\label{eq:sod_A_Q_final}
&\frac{\partial^2 \left\{ f(A(\gamma_{\phi=1}))\right\}}{\partial Q^2}
\notag\\
&~~~=\underbrace{\frac{\partial^2 \left\{ f(A(\gamma_{\phi=1}))\right\} }{\partial \{A^2(\gamma_{\phi=1})\}}}_{>0}
\underbrace{\left\{\frac{\partial\{A(\gamma_{\phi=1})\} }{\partial \gamma_{\phi=1}}\right\}^{2}}_{>0}
\underbrace{\left\{\frac{\partial \gamma_{\phi=1}}{\partial Q}\right\}^{2}}_{>0}\notag\\
&~~~~~~+\underbrace{\frac{\partial \left\{ f(A(\gamma_{\phi=1}))\right\} }{\partial \{A(\gamma_{\phi=1})\}}}_{<0}
\underbrace{\frac{\partial^2  \{A(\gamma_{\phi=1})\} }{\partial \gamma_{\phi=1}^2}}_{<0}
\underbrace{\left\{\frac{\partial \gamma_{\phi=1}}{\partial Q}\right\}^{2}}_{>0}\notag\\
&~~~~~~+\underbrace{\frac{\partial \left\{ f(A(\gamma_{\phi=1}))\right\} }{\partial \{A(\gamma_{\phi=1})\}}}_{<0}
\underbrace{\frac{\partial  \{A(\gamma_{\phi=1})\} }{\partial \gamma_{\phi=1}}}_{>0}
\underbrace{\frac{\partial^2  \gamma_{\phi=1} }{\partial Q^2}}_{=0}>0.
\end{align}
Then, following \eqref{eq:sod_A_Q_final}, we find that $\frac{\partial^2 \{\epsilon(Q)\}}{\partial Q^2}>0$ for $Q>Q_0$, due to $\frac{\partial^2 \left\{ f(A(\gamma_{\phi=1}))\right\}}{\partial Q^2}>0$.

Based on the analysis above, the sign of $\frac{\partial^2 \{P_{\epsilon}(Q)\} }{\partial Q^2}$ in \eqref{sod:P_epsilon} is determined as
\begin{align}
\frac{\partial^2 \{P^{\phi=1}_{\epsilon}(Q)\} }{\partial Q^2}
&\!=\!\!\!\!\underbrace{\frac{\partial^2 \{p_t(Q)\}}{\partial Q^2}}_{< 0~\text{for}~0 < Q < P_{\mathrm{max}}(N_{\mathrm{t}}-1)}
\underbrace{\!\!\biggl(\epsilon(Q)-1\biggr)}_{\leq 0}
\notag\\
&~~~~~~~~
+ 2 \underbrace{\frac{\partial \{p_t(Q)\}}{\partial Q}}_{< 0}
\underbrace{\frac{\partial \{\epsilon(Q)\}}{\partial Q}}_{< 0}
\notag\\
&~~~~~~~~
+ \underbrace{p_t(Q)}_{> 0}
\underbrace{ \frac{\partial^2 \{\epsilon(Q)\}}{\partial Q^2}}_{> 0~\text{for}~Q>Q_0}~\!\!\!\!\!>\!0.
\end{align}
%%%%%%%%%
To summarize, we prove that $\frac{\partial^2 \{P^{\phi=1}_{\epsilon}(Q)\} }{\partial Q^2}>0$ for $Q_0<Q< P_{\mathrm{max}}(N_{\mathrm{t}}-1)$, which completes the proof.

\bibliographystyle{IEEEtran}
\bibliography{Mybib_CIPC}

\end{document}